%% file: main-icse.tex
\documentclass[conference]{IEEEtran}
\IEEEoverridecommandlockouts

\input{metasection-imports}
\usepackage{orcidlink} 

\usepackage{cite}
\usepackage{amsmath,amssymb,amsfonts}
\usepackage{algorithmic}
\usepackage{graphicx}
\usepackage{textcomp}
\usepackage{xcolor}
\def\BibTeX{{\rm B\kern-.05em{\sc i\kern-.025em b}\kern-.08em
    T\kern-.1667em\lower.7ex\hbox{E}\kern-.125emX}}

\newif\ifblindreview

\blindreviewfalse  

\ifblindreview
  \newcommand{\anon}[1]{{\color{orange}ANONYMIZED}}
\else
  \newcommand{\anon}[1]{#1}
  
\fi

\begin{document}

\title{Are We Lost in the Woods? Detecting Silent Semantic Faults for Random Forest Classifiers with Data-informed Static Analysis
\thanks{This work was partially supported by the Wallenberg AI, Auto\-nomous Systems and Software Program~(WASP) funded by the Knut and Alice Wallenberg Foundation, and was in collaboration with Software Center Project~61 and Vinnova CoDig competence center. We thank the WARA-OPS for their computing resources.}}

\ifblindreview
\author{\IEEEauthorblockN{\textit{Anonymous Authors}}}
\else
\author{\IEEEauthorblockN{Willem Meijer~\orcidlink{0000-0001-8482-3917}}
\IEEEauthorblockA{\textit{Linköping University}\\
Linköping, Sweden \\
willem.meijer@liu.se
}
\and
\IEEEauthorblockN{Louis Ohl~\orcidlink{0000-0002-3467-5167}}
\IEEEauthorblockA{\textit{Linköping University}\\
Linköping, Sweden \\
louis.ohl@liu.se
}
\and
\IEEEauthorblockN{Kristian Sandahl~\orcidlink{0000-0002-3052-5604}}
\IEEEauthorblockA{\textit{Linköping University}\\
Linköping, Sweden \\
kristian.sandahl@liu.se
}
\and
\IEEEauthorblockN{Dániel Varró~\orcidlink{0000-0002-8790-252X}}
\IEEEauthorblockA{\textit{Linköping University}\\
Linköping, Sweden \\
daniel.varro@liu.se
}
}
\fi

\maketitle

\begin{abstract}
\input{section-abstract}

\end{abstract}


\input{metasection-commands}

\input{metasection-code-formatting-python}

\input{metasection-code-formatting-json}
\input{metasection-takeaway}

\input{section-introduction}
\input{section-background}

\input{section-related-work}

\input{section-implementation}
\input{section-evaluation}
\input{section-results}
\input{section-threats}
\input{section-discussion}

\input{section-conclusion}

\input{metasection-data-availability}

\input{metasection-credits}

\bibliographystyle{IEEEtranN}
\bibliography{references-doi2bib}

\end{document}

%% file: metasection-imports.tex
\usepackage[numbers,square,sort&compress]{natbib}
\usepackage[hidelinks]{hyperref}

\usepackage{tabularx}
\usepackage{subcaption}
\usepackage{lipsum}

\usepackage{enumitem}
\usepackage{textcomp}

\usepackage{listings}    
\usepackage{xcolor}      
\usepackage{float}       
\usepackage{newfloat}    

\usepackage[most]{tcolorbox}
\usepackage{xcolor}

\usepackage{cleveref}

\usepackage[acronym]{glossaries-extra}

%% file: section-abstract.tex
While machine learning~(ML) software necessitates effective quality assurance,  ML engineers still encounter silent semantic faults, such as imbalanced datasets, that degrade prediction performance without apparent symptoms.
These faults are typically detected after expensive training cycles, causing significant resource waste.

We propose a data-informed static analysis technique to detect silent semantic faults in ML scripts that use the popular random forest classifier.
Our approach extracts ML pipelines into directed acyclic graphs and evaluates them against formalized API contracts to detect structural, data, and hyperparameter faults.
Our analysis uses aggregated data properties, enabling fault detection even when datasets are inaccessible due to confidentiality restrictions.

We implemented this technique in an open-source tool, dille, and evaluated it on real-world Kaggle notebooks that use the random forest classifier.
Our results demonstrate that the tool identifies relevant semantic faults with~91\% precision and sub-second runtime overhead, making it suitable for integration into integrated development environments, agentic workflows, and continuous integration pipelines.
Our empirical study reveals that~12\% to~18\% of existing ML notebooks that use the random forest classifier are affected by silent semantic faults, highlighting the immediate practical utility of data-informed static analysis in reducing the burden of ML debugging.

%% file: metasection-commands.tex
\newfloat{lstfloat}{htbp}{lop}
\floatname{lstfloat}{Listing}
\def\lstfloatautorefname{Listing} 

\newcommand{\qt}[1]{\textit{``#1''}}

\newcommand{\myquestion}[2]{\begin{enumerate}[leftmargin=1cm]\item[\textbf{#1}]\textit{#2}\end{enumerate}}
\newcommand{\rqn}[1]{RQ\textsubscript{#1}}

\newcolumntype{Y}{>{\centering\arraybackslash}X}

\newcommand{\secref}[1]{\hyperref[#1]{Section}~\ref{#1}}

\crefformat{section}{\S#2#1#3}
\crefformat{subsection}{\S#2#1#3}
\crefformat{subsubsection}{\S#2#1#3}
\crefrangeformat{section}{\S\S#3#1#4 to~#5#2#6}
\crefmultiformat{section}{\S\S#2#1#3}{ and~#2#1#3}{, #2#1#3}{ and~#2#1#3}

\renewcommand{\subsectionautorefname}{Section}

%% file: metasection-code-formatting-python.tex
\definecolor{background}{HTML}{EEEEEE}
\definecolor{deepblue}{rgb}{0,0,0.5}
\definecolor{deepred}{rgb}{0.6,0,0}
\definecolor{deepgreen}{rgb}{0,0.5,0}
\definecolor{lightgray}{rgb}{0.95,0.95,0.95}
\definecolor{red}{rgb}{1,0,0}
\definecolor{sand}{rgb}{0.65, 0.48, 0.02}
\definecolor{lightblue}{rgb}{0.02, 0.58, 0.65}
\definecolor{steelblue}{RGB}{70, 130, 180}

\DeclareFixedFont{\ttb}{T1}{txtt}{bx}{n}{12} 
\DeclareFixedFont{\ttm}{T1}{txtt}{m}{n}{12}  

\newcommand\pythonstyle{\lstset{
    language=Python,
    basicstyle=\footnotesize\ttfamily,
    morekeywords={self},                    
    keywordstyle=\footnotesize\color{deepblue},
    emph={__init__,np,sklearn,dille,nbconvert,linear_model,LogisticRegression,numpy,pandas,torch,pd,preprocessing,StandardScaler,DataFrame,random,GaussianNB,naive_bayes,svm,SVC,RandomForestClassifier,LabelEncoder,matplotlib,plt,sns,seaborn,DecisionTreeClassifier,KMeans,PCA,MLPClassifier,KNeighborsClassifier,GradientBoostingClassifier,XGBClassifier,Ridge,Lasso,Pipeline,dille,
    ExtraTreesClassifier,AdaBoostClassifier,BaggingClassifier,VotingClassifier,StackingClassifier,ElasticNet,SVR,LinearSVC,NuSVC,XGBRegressor,LGBMClassifier,LGBMRegressor,CatBoostClassifier,DBSCAN,AgglomerativeClustering,TSNE,IsolationForest,
    MinMaxScaler,RobustScaler,Normalizer,OneHotEncoder,OrdinalEncoder,SimpleImputer,ColumnTransformer,PolynomialFeatures,SelectKBest,RFE,GridSearchCV,RandomizedSearchCV,KFold,StratifiedKFold,
    nn,optim,DataLoader,Dataset,Sequential,Conv2d,Linear,ReLU,Softmax,Sigmoid,Adam,SGD,Tensor,Module,Parameter,functional,F,autograd,keras,layers,models,callbacks,TensorFlow,tf,
    scipy,statsmodels,sm,nltk,spacy,cv2,PIL,Image,plotly,go,px,json,math,os,sys,time,datetime}, 
    emphstyle=\footnotesize\color{deepred},
    emph={[2]model,X,y,data,columns,df,scaler,scaled_data,scaled_df,gnb,svc,max_samples,criterion,n_estimators,bootstrap,random_seed,X_train,X_test,y_train,y_test,preds,y_pred,clf,params,grid_search,random_state,test_size,cv,n_jobs,verbose,history,numeric,
    train_df,test_df,val_df,X_val,y_val,features,target,labels,batch,images,tensors,inputs,outputs,input_dim,output_dim,input_shape,vld,pred,enc,
    learning_rate,lr,batch_size,epochs,max_depth,min_samples_split,min_samples_leaf,subsample,colsample_bytree,gamma,alpha,penalty,solver,momentum,weight_decay,dropout,hidden_size,hidden_dim,n_clusters,k,leaf_size,
    regressor,classifier,pipeline,search,gs,rs,encoder,imputer,poly,explainer,optimizer,loss_fn,scheduler,device,rf,
    accuracy,loss,val_loss,score,metrics,cm,cr,fpr,tpr,auc,roc_auc,f1,precision,recall,probs,y_score,
    seed,config,path,file_path,mapping,classes_},                  
    emphstyle={[2]\footnotesize\color{lightblue}},
    emph={[3]array,fit,nan,rand,fit_transform,transform,read_csv,predict,predict_proba,head,tail,describe,info,drop,groupby,apply,fillna,dropna,concat,merge,plot,show,reshape,append,values,tolist,train_test_split,evaluate,score,
    set_index,reset_index,unique,value_counts,sort_values,pivot_table,melt,rename,to_csv,to_datetime,to_numeric,isin,query,duplicated,iloc,loc,astype,sample,
    linspace,arange,zeros,ones,mean,std,sum,min,max,argmax,argmin,concatenate,stack,transpose,flatten,abs,log,exp,sqrt,
    fit_predict,decision_function,get_params,set_params,inverse_transform,get_feature_names_out,cross_val_score,partial_fit,compile,summary,step,zero_grad,backward,item,save,load,
    legend,xlabel,ylabel,title,subplot,subplots,savefig,scatter,hist,boxplot,heatmap,pairplot,tight_layout,grid,xticks,yticks,
    print,range,len,enumerate,zip,sorted,map,filter,open},                  
    emphstyle={[3]\footnotesize\color{sand}},
    emph={[4]True,False,None},
    emphstyle={[4]\footnotesize\color{steelblue}},
    stringstyle=\color{deepgreen},
    frame=tb,                               
    showstringspaces=false,
    numbers=left,                          
    numberstyle=\scriptsize\color{red},    
    stepnumber=1,                          
    numbersep=4pt,                         
    backgroundcolor=\color{lightgray},     
    xleftmargin=1.5em,                       
    captionpos=b,
    commentstyle=\color{deepgreen}, 
    morecomment=[l]{\#},            
}}
            
\lstnewenvironment{python}[1][]
{
\pythonstyle
\lstset{#1}
}
{}

\DeclareRobustCommand\pythoninline[1]{{\normalsize\pythonstyle\lstinline!#1!}}

%% file: metasection-code-formatting-json.tex
\definecolor{background}{HTML}{EEEEEE}

\definecolor{deepblue}{rgb}{0,0,0.5}
\definecolor{deepred}{rgb}{0.6,0,0}
\definecolor{deepgreen}{rgb}{0,0.5,0}
\definecolor{lightgray}{rgb}{0.95,0.95,0.95}
\definecolor{red}{rgb}{1,0,0}

\DeclareFixedFont{\ttb}{T1}{txtt}{bx}{n}{12} 
\DeclareFixedFont{\ttm}{T1}{txtt}{m}{n}{12}  

\newcommand\jsonstyle{\lstset{
    language={},
    basicstyle=\footnotesize\ttfamily,
    frame=tb,
    showstringspaces=false,
    numbers=left,
    numberstyle=\scriptsize\color{red},
    stepnumber=1,
    numbersep=4pt,
    backgroundcolor=\color{lightgray},
    xleftmargin=1.5em,
    captionpos=b,
    morekeywords={true,false,null},
    keywordstyle=\footnotesize\color{deepred},
    morestring=[b]",
    stringstyle=\color{deepgreen}, 
    moredelim=**[is][\color{deepblue}]{@key}{@}, 
    moredelim=**[is][\color{deepgreen}]{@val}{@}, 
    literate=
       {\{}{{{\color{black}\{}}}1
       {\}}{{{\color{black}\}}}}1
       {[}{{{\color{black}[}}}1
       {]}{{{\color{black}]}}}1
       {:}{{{\color{black}:}}}1
       {,}{{{\color{black},}}}1
}}

\lstnewenvironment{json}[1][]
{
  \jsonstyle
  \lstset{#1}
}
{}

\DeclareRobustCommand\jsoninline[1]{{\jsonstyle\lstinline!#1!}}

%% file: metasection-takeaway.tex
\definecolor{steelblue}{RGB}{70, 130, 180} 
\definecolor{darksteelblue}{RGB}{34, 62, 88} 
\definecolor{lightgray}{RGB}{242, 242, 242} 

\newtcolorbox[auto counter, number format=\Alph]{takeawaybox}[1][]{
    sharp corners,
    colback=lightgray,
    colframe=steelblue,
    leftrule=3pt,
    rightrule=0pt,
    toprule=0pt,
    bottomrule=0pt,
    enhanced,
    fonttitle=\bfseries\sffamily,
    coltitle=steelblue,
    title={}, 
    attach title to upper,
    after title={},
    boxsep=5pt,
    arc=0pt
}

\newcommand{\takeaway}[2]{%
    \begin{takeawaybox}
        {\color{darksteelblue} \textbf{#1}:}~#2
    \end{takeawaybox}
}

%% file: section-introduction.tex
\section{Introduction}

Due to the increasing adoption of machine learning (ML)-based software systems~\citep{oecdbcginsead_adoption_2025, arroyabe_analyzing_2024}, the need for adequate software quality assurance techniques for such systems is also growing continuously~\citep{shivashankar_maintainability_2022, santhanam_quality_2020, bogner_characterizing_2021, cote_quality_2024}.
Existing research has delivered best practices for ML development~\citep{serban_software_2024}, and various techniques for data validation and testing~\citep{whang_data_2023, kumarOpportunitiesChallengesDataCentric2024, albelali_testing_2025} or for revealing architectural issues~\citep{nazir_architecting_2024, bucaioni_checklist_2025}. 

ML experts still often struggle to detect \textit{silent semantic faults} caused by the inappropriate use of ML models~\citep{meijer_data-aware_2026, humbatova_taxonomy_2020, de_santana_bug_2024, khairunnesa_what_2023}.
For example, one may easily forget about correlating features before training a correlation-sensitive model.
But such silent semantic faults often lack apparent side-effects (like a crash or incorrect output).
As such, ML experts may only suspect their existence from poor prediction performance, or excessive compute resource usage.
To reveal such faults after training a model, one must (iteratively) investigate the training results, determine if suboptimal results are due to a bug or just the wrong choice of model, locate the bug in the pipeline and/or data, fix the bug, and re-train the model.
Therefore, detecting and debugging semantic faults in ML scripts is a highly time- and resource-consuming process~\citep{lai_comparative_2024, morovati_bug_2024}.



\paragraph*{Problem statement}
Existing solutions to detect silent semantic faults in ML scripts may identify incorrect hyperparameters and certain training issues~\citep{gao_refty_2022, reimann_safe-ds_2023, shivashankar_mlscent_2025, ahmed_design_2023}.
However, they also have major conceptual limitations: they may not consider data at all~\citep{shivashankar_mlscent_2025, ahmed_design_2023, turcotte_fault_2025}, or only consider structural properties~\citep{gao_refty_2022, reimann_safe-ds_2023}, missing semantic properties such as distributions or correlations.

A recent new ideas paper \citep{meijer_data-aware_2026} proposed a data-aware  
approach to detect semantic faults statically in ML code. Static detection before training a model allows ML developers to detect faults while writing their code, thus reducing the time-consuming manual investigation of the results.
The paper conducted a preliminary analysis of 21 notebooks using the Scikit-learn library and detected semantic faults in 5 notebooks.

The current paper provides the first in-depth investigation of data-informed static analysis of ML scripts.
We have decided to focus on an in-depth study in the context of random forest classifiers.
Random forest~\citep{breiman_random_2001} represents a simple but very popular ML classifier used by data scientists in a multitude of application domains~\citep{fox_assessing_2017, cappelli_random_2024} and outperforms neural networks on problems that use tabular data~\citep{grinsztajn_why_2022}.
Therefore, if our data-informed static fault detection is successful for random forest, it has immediate practical benefits. Moreover, some of the contracts are transferable to other ML algorithms.




\quad 

\paragraph*{Objectives}
    In this paper, we propose a novel data-informed static analysis technique to detect silent semantic faults in ML scripts that use the random forest classifier.
    After formalizing silent semantic faults as API contracts for random forests and calculating data properties for datasets, we extract ML pipelines from scripts in the form of directed acyclic graphs~(DAGs).
    We calculate pipeline-specific slices of the DAG to identify how data transformations affect data properties and reveal silent semantic faults as violations.
    As a key conceptual extension over \citep{meijer_data-aware_2026}, our (data-informed) analysis can operate with only aggregated data properties, thus, we can detect semantic faults without accessing confidential datasets. The specific contributions of the paper are as follows:
    \begin{itemize}
        
        \item We extract ML pipelines from scripts into an abstract graph format, which captures training steps, data transformations, and model hyperparameters.

        \item We detect silent semantic faults by evaluating API contracts on pipeline-specific slices of the graph.
        
        \item We provide an open-source prototype tool \pythoninline{dille}~\citep{meijer_experiment_2026} implementing our analysis technique.
        
        \item We provide an experimental evaluation of our tool to assess the frequency, precision, and relevance of silent semantic fault detection.
        \item We carry out an empirical study on Kaggle notebooks that use the random forest classifier, which reveals that a large portion of notebooks are affected by silent semantic faults.
    \end{itemize}

\paragraph*{Added value}
    Our tool can detect a unique class of relevant silent semantic faults in ML scripts using random forest classifiers with high precision~($91\%$).
    With its sub-second runtime, our tool can be integrated into development environments, agentic workflows, and continuous integration/delivery~(CI/CD) pipelines.
    Our analysis is effective even when relying on precomputed data properties instead of the real dataset, which is beneficial when strict data confidentiality restrictions are in effect.

%% file: section-background.tex
\input{asset-table-semantic-faults}

\section{Background}

    The following sections introduce the random forest classifier and the concept of silent semantic faults. 

\subsection{The Random Forest Classifier}
\label{sec:random-forest}
    
    The random forest classifier~\citep{breiman_random_2001, sklearn-rf-doc} is a simple yet popular ML algorithm used for classification.
    It is an ensemble learning method that predicts the class of data points by taking the majority vote of multiple decision tree sub-classifiers~\citep{breiman_classification_2017, sklearn-dt-doc}.
    Strengths of this algorithm are that more trees reduce variance without increasing bias, its ability to capture non-linear relationships, and its good performance when the number of data features is far greater than the number of data points.
    In addition, it can outperform neural networks when analyzing tabular data~\citep{grinsztajn_why_2022}, and can be effectively used in diverse tasks such as predicting water condition~\citep{fox_assessing_2017} or identifying primary factors of cardiovascular and respiratory disease~\citep{cappelli_random_2024}.

    Training a random forest classifier is equivalent to training multiple decision trees.
    These trees are trained on a random subset of the training data points and features to minimize \textit{correlation between trees} while maximizing the information available for predictions.
    This makes the model resilient to minor data changes and prevents it from becoming dependent on particular features, lowering the chance that multiple trees make the same incorrect predictions.

    Decision trees are prone to correlation because they are constructed by greedily adding decision points based on their \qt{informativeness.}
    By iterating through all features and potential decision points, the algorithm selects the optimal point to separate predicted classes.
    This ends when the tree has a specific size or a new split yields insufficient improvement.


    \paragraph*{Example}
        \autoref{lst:running-example} illustrates a short yet typical use of the random forest classifier, implementing an insurance claim prediction pipeline~\citep{wiryaseputra2023health}.
        The code imports a training dataset~(line~1) and separates the outcome feature \pythoninline{"Response"} from the predictor data~(lines~2 and~3).
        It then initializes a \pythoninline{LabelEncoder}~(line~4) and uses this to transform the categorical feature \pythoninline{"Damage"}~(indicating whether the claimant's vehicle was damaged) to a number~(line~5).
        After splitting the data for training and testing~(lines~6 and~7), a \pythoninline{RandomForestClassifier} is constructed~(line~8) and trained~(line~9).
        Finally, a separate validation set is loaded~(line~10), transformed in the same way as the training data~(lines~11 and~12), and the model's prediction performance is validated~(line~13).

\input{asset-listing-running-example}

\subsection{Silent Semantic Faults}
\label{sec:semantic-faults}
    
    Based on ISO/IEC/IEEE~24765~\citep{iso24765}, a silent semantic fault in ML code is \qt{a mistake~(a fault) in training code involving a violation of a ML algorithm's intended use~(its semantics) that could cause it to perform outside specified limits~(e.g., biased predictions), without causing it to crash~(it is silent).}
    In our context, this refers to faults caused by misusing the random forest algorithm.
    Silent semantic faults are difficult to detect due to the lack of clearly erroneous symptoms like a crash or unreasonable results.

    \autoref{tab:fault-taxonomy} provides an overview of the 11 silent semantic faults specific to the random forest classifier considered in this article, which we identified by surveying literature and analyzing the code and documentation of Scikit-learn~\citep{pedregosa_scikit-learn_2011, sklearn-rf-doc}.
    Code analysis is particularly crucial as implementations commonly deviate from their theoretical counterparts.
    This might add implementation-specific fault types, while removing theoretical ones.
    
    We define three main types of semantic faults.
    (1)~\emph{Hyperparameter faults} are the simplest type, indicating incorrectly specified model hyperparameters.
    (2)~\emph{Data faults} describe mismatches between ML algorithms and the datasets used to train them.
    Although the origins of these faults can typically be identified by inspecting the dataset, they do not always cause a problem, as various data preprocessing steps can mitigate them.
    Finally, (3)~\emph{structural faults} represent the incorrect integration of pipeline steps.
    These can negatively affect the trained model's performance, even though there are no issues with the dataset or algorithms themselves.

    \paragraph*{Example}
        While \autoref{lst:running-example} shows a typical example of using a random forest classifier, it also contains a silent semantic fault:~unequal data preprocessing~(S2) caused by incorrect use of the \pythoninline{LabelEncoder}.
        This model enumerates categorical values in the dataset based on order of appearance.
        Consequently, because this model is first fitted on the training data~(line~5) and then refitted on the validation data~(line~11), values with the same categorical values acquire different encodings when their order of appearance is different.
        While \pythoninline{"Yes"} was encoded as~1 in the training data, it was encoded as~0 in the validation data, negatively affecting the model's prediction performance.
        This fault is resolved by using the \pythoninline{transform} method to transform the validation data~(line~11), ensuring the encoding is identical.

%% file: asset-table-semantic-faults.tex
\begin{table*}[bt]
    \centering
    \caption{Overview of semantic faults, differentiating between hyperparameter~(H), data~(D), and structural~(S) faults}
    \label{tab:fault-taxonomy}
    \begin{tabularx}{\textwidth}{|p{2cm}|X|}
        \hline
        \textbf{Fault Name} & \textbf{Description} \\
        \hline\hline

        
        \textit{Missing Random Seed~(H0)} & To ensure experiment reproducibility, random forest requires a fixed random seed. Omitting the \pythoninline{random_state} parameter in the \pythoninline{RandomForestClassifier} causes inconsistent model behavior across different executions, hindering the reliable validation and comparison of experimental results~\citep{sklearn-rf-doc}. \\

        \textit{Invalid hyperparameters~(H1)} & Scikit-learn algorithms impose specific constraints and best practices on hyperparameters. For instance, \pythoninline{RandomForestClassifier} restricts the \pythoninline{criterion} to a specific list of strings~\citep{sklearn-rf-doc}. \\

    
        \textit{Diminishing returns in huge forests~(H2)} & While random forests do not overfit as the number of trees increases~\citep{breiman_random_2001}, performance gains typically converge asymptotically after the first 100 trees~\citep{probst_hyperparameters_2019} while they do continue to decrease runtime efficiency. \\

        \hline\hline



        \textit{Class imbalance~(D1)} & This fault occurs when target classes are not equally represented in a dataset. This causes model bias and renders performance metrics unreliable, as the model favors the majority class~\citep{weiss_mining_2010, cieslak_learning_2008, chen_using_2004}. Consequently, the minority class is more likely to be misclassified. This can be managed, e.g., by assigning weights to data points. \\

    
        \textit{Correlated features~(D2)} & Strongly correlated features undermine the independence between trees~\citep{breiman_random_2001, nicodemus_behaviour_2010, efron_prediction_2020}. Even though trees are trained on a random feature subset, the greedy feature selection algorithm may use them interchangeably. This often decreases the calculated importance of both features, hindering the interpretability of model predictions. \\

        \textit{Unconstrained trees~(D3)} & Training trees on large datasets without constraints often results in excessive model growth, decreasing performance efficiency while increasing the risk of overfitting. To mitigate this, it is recommended to tune structural hyperparameters (as discussed in sklearn issue \href{https://github.com/scikit-learn/scikit-learn/issues/8594}{\#8594} and pull request \href{https://github.com/scikit-learn/scikit-learn/pull/8721}{\#8721}). \\

        \textit{Unequal row count~(D4)} & The number of observations in the predictor matrix $X$ and the target vector $y$ must remain identical. Discrepancies between these row counts prevent proper sample alignment, typically resulting in execution failures during the model fitting process. \\

        \textit{Disproportionate random feature count~(D5)} & When the number of features used to train trees is very low with respect to the number of features in the dataset, the forest's performance typically drops due to the lack of information inside trees~\citep{probst_hyperparameters_2019}. \\

        \textit{Overlapping training samples~(D6)} & Training trees on larger random subsets of the dataset increases the likelihood that different trees share identical data points. While providing more training data per tree, this overlap increases inter-tree correlation, which can negatively affect the ensemble's overall classification performance~\citep{probst_hyperparameters_2019}. \\

        
        \hline\hline

        \textit{Non-random train and test data~(S1)} & Training aims to maximize performance on novel data by evaluating the model on a separate test set. Splitting data non-randomly (e.g., by index) can cause underperformance if the data order carries meaning~\citep{efron_prediction_2020}. For example, the equipment used to collect the data points. \\


        \textit{Unequal data preprocessing~(S2)} & Accurate evaluation requires identical preprocessing of train and test data, as inconsistent application makes it impossible to attribute performance to the model or preprocessing. For example, separately applying \pythoninline{LabelEncoder}~\citep{sklearn-le-doc} to the train and test data can yield different encodings for the same input values. \\
    

         \hline
    \end{tabularx}
\end{table*}

%% file: asset-listing-running-example.tex
\begin{lstfloat}[!b]
\begin{python}[caption={An example based on a real notebook~\citep{wiryaseputra2023health} of a random forest classifier script that contains a silent semantic fault:~unequal data preprocessing~(S2).}, label={lst:running-example}]
df = pd.read_csv("insurance_train.csv")
X = df.drop(columns=["Response"])
y = df["Response"]
enc = preprocessing.LabelEncoder()
X["Damage"] = enc.fit_transform(X["Damage"])
X_train, X_test, y_train, y_test =
    train_test_split(X, y, random_state=42)
rf = RandomForestClassifier(random_state=42)
rf.fit(X_train, y_train)
vld = pd.read_csv("insurance_valid.csv")
vld["Damage"] = enc.fit_transform(vld["Damage"])
vld = vld.drop(columns=["Response"])
pred = rf.predict(vld)
\end{python}
\end{lstfloat}

%% file: section-related-work.tex
\section{Related Work}
\label{sec:related-work}

    Code analysis of ML~\citep{gao_refty_2022,ahmed_design_2023,reimann_safe-ds_2023,shivashankar_mlscent_2025, dolcettiPYRAHighlevelLinter2026} and statistical software~\citep{turcotte_fault_2025,turcotte_expressing_2025} can be divided based on two main dimensions:~1)~data-awareness, and~2)~analysis period.
    We distinguish between data-agnostic~\citep{shivashankar_mlscent_2025, turcotte_fault_2025, reimann_safe-ds_2023, gao_refty_2022, ahmed_design_2023} and data-aware techniques~\citep{reimann_safe-ds_2023, gao_refty_2022, turcotte_expressing_2025}, based on whether properties of the train and test datasets are actively used in analyses.
    We further split static~\citep{shivashankar_mlscent_2025, turcotte_fault_2025} from runtime analysis techniques~\citep{reimann_safe-ds_2023, gao_refty_2022, turcotte_expressing_2025, ahmed_design_2023}, highlighting when these tools detect faults:~before or during execution.

\paragraph*{Data-agnostic analysis}
    Various tools identify code faults and other coding problems by enforcing API constraints without inspecting data~\citep{ahmed_design_2023, gao_refty_2022, reimann_safe-ds_2023, shivashankar_mlscent_2025, shivashankar_mlscent_2025, turcotte_fault_2025}.
    Most commonly, these solutions detect missing or incompatible hyperparameters~\citep{ahmed_design_2023, gao_refty_2022, shivashankar_mlscent_2025, reimann_safe-ds_2023, dolcettiPYRAHighlevelLinter2026, hong_investigating_2024} or structural faults~\citep{dolcettiPYRAHighlevelLinter2026, hong_investigating_2024}.
    This can be extended using heuristics to identify data issues~\citep{shivashankar_mlscent_2025, turcotte_fault_2025}; for example, by assuming that a lack of visualization implies a failure to check for class imbalance~\citep{turcotte_fault_2025}.
    Finally, \citet{ahmed_design_2023} use runtime analysis to detect training-specific problems like slow convergence when training neural networks.

\paragraph*{Data-aware analysis}
    Some recent work incorporates data directly into the analysis~\citep{reimann_safe-ds_2023, gao_refty_2022, turcotte_expressing_2025, dolcettiPYRAHighlevelLinter2026}.
    Examples include verifying data-model structural compatibility~\citep{gao_refty_2022} or using domain-specific languages to ensure valid schema access and pipeline sequencing~\citep{reimann_safe-ds_2023}.
    \citet{dolcettiPYRAHighlevelLinter2026} transcend this by creating a data science-specific type system that loads datasets to infer data types, missing data, and duplicate data entries.
    Despite their utility, these methods do not detect semantic faults like class imbalances.
    To some extent, this is resolved by \citet{turcotte_expressing_2025}, who inject statistical tests at runtime to validate statistical model assumptions.

\paragraph*{Semantic faults in ML software}
    Various empirical studies characterize bugs in ML \citep{chen_towards_2025, wang_why_2025, de_santana_bug_2024, de_santana_bug_2024} and deep learning \citep{zhang_empirical_2018, islam_comprehensive_2019, zhang_empirical_2020, hong_investigating_2024, humbatova_taxonomy_2020, morovati_bug_2024}, which are sometimes addressed in the context of notebooks~\citep{chen_towards_2025, wang_why_2025, de_santana_bug_2024}.
    Key semantic faults include API misuse~\citep{zhang_empirical_2018, islam_comprehensive_2019, zhang_empirical_2020, humbatova_taxonomy_2020, chen_towards_2025, wang_why_2025, morovati_bug_2024}, misunderstandings about data~\citep{islam_comprehensive_2019, zhang_empirical_2020, chen_towards_2025, wang_why_2025}, and structural issues in the pipeline~\citep{hong_investigating_2024, humbatova_taxonomy_2020}.
    These categories provide a general frame for the specific semantic faults investigated in this work.

\paragraph*{Novelty}
    We present the first data-informed static analysis technique to detect silent semantic faults using only aggregate data properties, effectively removing the dependency on complete datasets.
    While existing static tools are largely data-agnostic, and data-aware techniques are typically restricted to detecting data-structural faults, we actively incorporate data properties to reveal algorithm-specific faults.
    Furthermore, whereas prior empirical studies categorize broad fault types such as API misuse or data confusion, our study provides a fine-grained investigation of \textit{random-forest pipelines}, uncovering semantic faults that generalized taxonomies miss.

%% file: section-implementation.tex
\section{Data-informed Static Analysis}
\label{sec:implementation}

    To detect semantic faults in ML code, we propose a data-informed static analysis technique that natively exploits aggregate statistics of datasets.
    We implement this procedure in our tool, called \pythoninline{dille}~\citep{meijer_experiment_2026}.
    Initial ideas of such analysis have been explored in a recent new ideas paper~\citep{meijer_data-aware_2026}, to which we add the key extension that our (data-informed) analysis operates with only aggregated data properties.
    This approach analyzes Python ML code by creating an abstract graph representation of the pipeline implemented in the code.
    This graph representation enables data- and control-flow analysis by capturing how data is loaded, preprocessed, and used to train ML models.
    In turn, semantic faults can be detected by encoding algorithm-specific prerequisites and assumptions as contracts and by verifying if the datasets, in combination with their preprocessing steps, are compliant.
    Data-informed static analysis contains three main steps: code canonicalization~(\autoref{sec:code-standardization}), pipeline extraction~(\autoref{sec:pipeline-extraction}), and fault detection~(\autoref{sec:fault-detection}).

    \subsection{Code Canonicalization}
    \label{sec:code-standardization}
        To ensure all scripts can be processed using the same analysis technique, the code of the ML script is canonicalized by a series of abstract syntax tree~(AST) transformations.
        First, we \emph{unfold calls to self-defined functions} by extracting their operations into the main body of the script.
        In many cases, such methods are used to create, e.g., a common preprocessing pipeline for different classifiers.
        We continue by \emph{replacing in-place assignments} with explicit assignments.
        For example, by replacing \pythoninline{df.drop(y, inplace=True)} with \pythoninline{df = df.drop(y)}.
        We then \emph{unfold for loops} that iterate over lists with explicitly defined contents.
        This commonly happens when a script does hyperparameter selection using grid-search, where all the considered options~(e.g., the number of trees in the forest) are defined as a list.
        This is followed by \textit{resolving library aliases}~(e.g., resolving the alias \pythoninline{pd} to \pythoninline{pandas}) to simplify API identification.
        Finally, we \textit{split nested and chained statements} into separate lines of code.
        These are very common in the scripts in our dataset~(see~\autoref{sec:data-collection}).
        For example, when feature selection and training are done on the same line.

\input{asset-listing-contract}

    \subsection{Pipeline Extraction}
    \label{sec:pipeline-extraction}
        ML scripts commonly implement multiple pipelines in parallel using different preprocessing steps and classifiers, which can be extracted as a directed acyclic graph~(DAG).

        After canonicalization, data transformations are explicitly captured as variable assignments~(statements like \pythoninline{X = enc.fit_transform(df)}).
        By iterating over all assignments in the AST, it becomes possible to identify the inputs and outputs of pipeline steps, where the output is equal to its left-hand side~(i.e., \pythoninline{X}) and the input is equal to the arguments passed to the function call on the right-hand side~(i.e., \pythoninline{df}).
        In turn, any input variable can be mapped to its most recent output variable, creating a DAG of the code's data/control flow, where each node is a pipeline step and each edge represents how the output of one step is used as the input of another.

        \paragraph*{Example}
            \autoref{fig:running-example-dag} shows the DAG representing a slice of \autoref{lst:running-example}.
            In contrast to the code, we can clearly see a training and a validation pipeline, which load a dataset from a \texttt{csv} file, remove the \pythoninline{"Response"} feature from the dataset, and encode the \pythoninline{"Damage"} column using the \pythoninline{LabelEncoder} without affecting the rest of the data in \pythoninline{df}.
            When this data is later used to train a model or make predictions, we know their exact preprocessing pipelines.

\input{asset-figure-running-example-dag-data}

\input{asset-figure-running-example-dag}

    \subsection{Fault Detection}
    \label{sec:fault-detection}
        To detect silent semantic faults, we define random-forest-specific API contracts that encode how the \pythoninline{RandomForestClassifier} APIs should be used, connecting the semantic faults in \autoref{tab:fault-taxonomy} to specific API endpoints.
        Beyond a name and description, each contract explicitly specifies the API and the specific endpoint to which it applies, and a Python method that evaluates whether a fault is made.
        For example, \autoref{lst:contract} shows the contract that links unequal data preprocessing~(S2) to the \pythoninline{predict} method.
        We established a suite of~$46$ API contracts that detect the semantic faults in \autoref{tab:fault-taxonomy}.

        Although pipeline extraction and fault detection steps are described separately, our implementation executes them in tandem.
        While traversing the AST to construct a DAG, the API contracts of the visited method calls are loaded, and their respective evaluator methods are used to enforce the contract using the current state of the DAG.

        \subsubsection{Data faults}
            The detection of data faults requires both the training pipeline and the dataset because pipelines might include steps that mitigate data faults.
            For example, although class imbalance~(D1) can be detected in the dataset by calculating the ratio between the majority and minority classes, this problem can be mitigated by assigning greater importance to the minority class during training or by oversampling it.

            Because our analysis is static~(i.e., it does not execute the analyzed code), it is impossible to compute data properties such as class imbalance on the literal data used to train a model.
            To resolve this issue, our solution extracts these statistics from the original dataset and tracks how the pipeline affects them.
            We do this using API guarantees, which, like API contracts, reference Python methods that encode how data properties are affected by preprocessing APIs.

            When data contracts are enforced, their corresponding evaluator method traverses the DAG backwards to identify what APIs were used to transform the data, loading their respective API guarantees.
            In turn, once the root of the DAG is reached~(i.e., where the dataset is loaded), the requested data property is calculated.
            This can be carried out in two ways, depending on our tool's configuration:~1)~extracting them from the dataset, or~2)~retrieving them from precomputed data properties.
            In turn, the acquired data property is transformed using the API guarantees, after which the contract can be evaluated.

             \paragraph*{Example}
                \autoref{fig:running-example-dag-data} shows the process to detect feature correlation~(D2).
                Starting at \pythoninline{fit}, the evaluator method first identifies relevant feature names by traversing backward toward the pipeline's root: \pythoninline{read_csv}.
                It retrieves the list of features from (precomputed) data, after which API guarantees are applied.
                The \pythoninline{drop} call is guaranteed to remove \pythoninline{"Response"}, while \pythoninline{fit_transform} and \pythoninline{train_test_split} have no effect.
                The evaluator then calculates pairwise feature correlations for all features except \pythoninline{"Response"} using the same process, registering a violation if any correlation exceeds a threshold.


        \subsubsection{Structural faults}
            Structural faults transcend individual datasets and algorithms, hence their detection relies heavily on the generated DAG.
            For example, the DAG can be used to detect whether two slices are structurally equivalent.

            \paragraph*{Example}
                To detect the unequal data preprocessing~(S2) in \autoref{fig:running-example-dag}, its corresponding contract is enforced on the \pythoninline{predict} call.
                Because this call specifies the used model, \pythoninline{rf}, the model's most recent \pythoninline{fit} call can be identified.
                Using the DAG, the specific input pipelines of \pythoninline{fit} and \pythoninline{predict} can be extracted and compared.
                Unequal preprocessing means two things:~1)~the type of preprocessing steps are equivalent~(e.g., if the \pythoninline{LabelEncoder} is used to encode one pipeline, it is also used to encode the other), and~2)~every \pythoninline{transform} or \pythoninline{fit_transform} call in the training pipeline has a matching \pythoninline{transform} call in the validation pipeline.
                In turn, traversing both pipelines simultaneously, both drop the \pythoninline{"Response"} column and use the \pythoninline{LabelEncoder} to transform the \pythoninline{"Damage"} column, indicating that the types of preprocessing steps are equivalent.
                However, the \pythoninline{fit_transform} in the training pipeline has no corresponding \pythoninline{transform} in the prediction pipeline, thus violating the second requirement.

        \subsubsection{Hyperparameter faults}
            Hyperparameter faults are the easiest fault type to detect, as hyperparameters are typically defined in code and independent of the pipeline.
            This enables their detection without considering the rest of the training pipeline.
            For example, a missing random seed~(H0) can be detected by checking whether the \pythoninline{random_seed} was set, and diminishing returns in huge forests~(H2) can be detected by reading \pythoninline{n_estimators} and comparing it to a threshold.

%% file: asset-listing-contract.tex
\begin{lstfloat}[!b]
\begin{json}[caption={The API contract that encodes unequal data preprocessing~(S2), containing the targeted API and endpoint, a description, and the evaluation method that enforces it.}, label={lst:contract}]
"name": "Unequal Data Preprocessing (S2)",
"description": "The train and test pipelines
       should be implemented identically...",
"api":"sklearn.ensemble.RandomForestClassifier",
"endpoint": "predict",
"eval_func": "evaluators.identical_pipelines"
\end{json}
\end{lstfloat}

%% file: asset-figure-running-example-dag-data.tex
\begin{figure}[b]
    \centering
    \includegraphics[width=0.838\columnwidth]{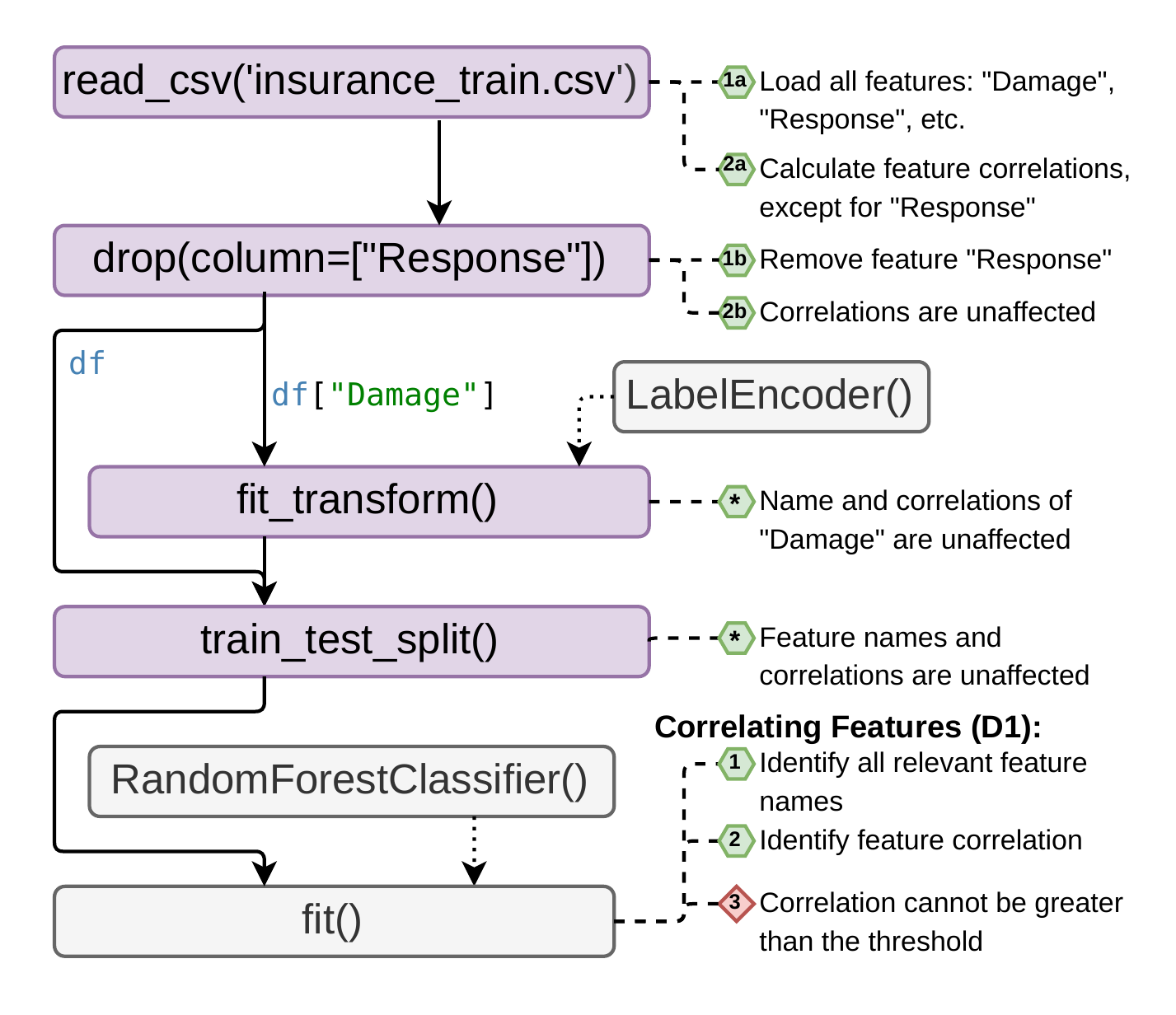}
    \caption{The DAG corresponding to a slice of \autoref{lst:running-example} showing the training pipeline~(purple nodes), highlighting the API contract requirements~(red diamond), and the evaluation steps~(green hexagons) used to detect correlating features~(D1).}
    \label{fig:running-example-dag-data}
\end{figure}

%% file: asset-figure-running-example-dag.tex
\begin{figure}[b]
    \centering
    \includegraphics[width=\columnwidth]{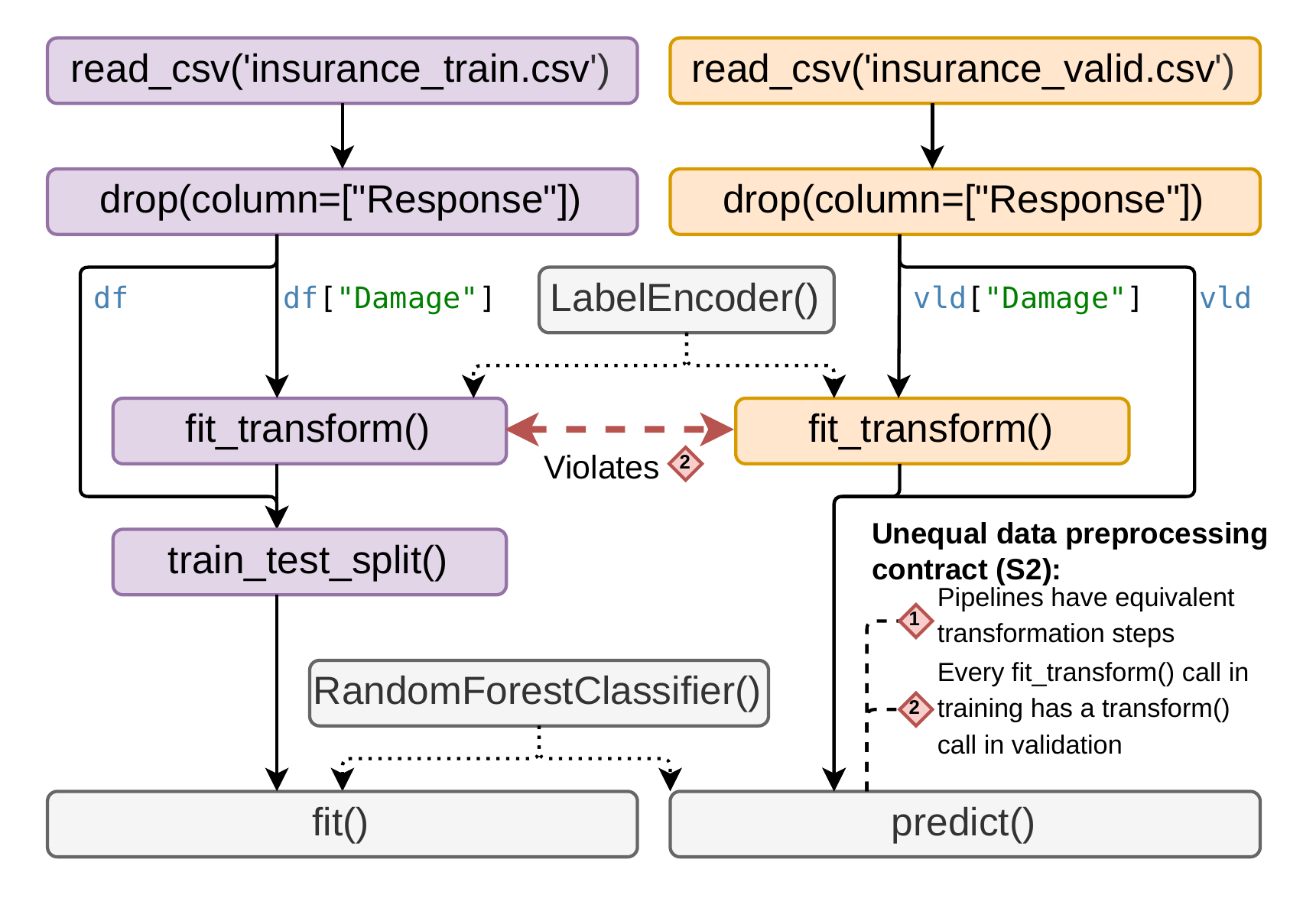}
    \caption{The DAG corresponding to a slice of \autoref{lst:running-example}, showing the training~(purple; left) and validation pipelines~(orange; right), and highlighting the API contract requirements~(red diamonds) used to detect unequal data preprocessing~(S2).}
    \label{fig:running-example-dag}
\end{figure}

%% file: section-evaluation.tex
\section{Evaluation}

    We evaluate data-informed static analysis in the context of Scikit-learn's \pythoninline{RandomForestClassifier}~\citep{pedregosa_scikit-learn_2011, sklearn-rf-doc} to address three research questions.  

    We evaluate the effectiveness of our analysis by focusing on three key dimensions of static analysis techniques~\citep{johnson_why_2013}:~the volume of reported issues~(\textit{frequency}), the correctness of these reports~(\textit{precision}), and the perceived utility of the identified faults~(\textit{relevance}).
    
    \myquestion{\rqn{1}}{How effective is data-informed static analysis to detect silent semantic faults in machine learning scripts?}

    Running static analysis causes additional runtime overhead.
    However, static analysis executed in various contexts has different time constraints. For example, while analysis in a \mbox{CI/CD} pipeline is less time-constrained~(minutes to hours), running an analysis in the background of an IDE requires very short runtimes~(sub-second).
    Therefore, we evaluate the runtime overhead of our procedure.

        
    \myquestion{\rqn{2}}{What is the runtime overhead of data-informed static analysis?}

    Based on the results of \rqn{1}, our tool can be used to empirically estimate a lower bound of the prevalence of semantic faults in ML training code that use random forest classifiers.
    While various prior studies categorize broad fault types such as API misuse~\citep{zhang_empirical_2018, islam_comprehensive_2019, zhang_empirical_2020, humbatova_taxonomy_2020, chen_towards_2025, wang_why_2025, morovati_bug_2024} and data confusion~\citep{islam_comprehensive_2019, zhang_empirical_2020, chen_towards_2025, wang_why_2025}, algorithm-specific silent semantic faults have not been studied.
    Because silent faults lie at the core of our solution, we perform this investigation.
    
    \myquestion{\rqn{3}}{How prevalent are silent semantic faults in machine learning notebooks?}

\subsection{Data Collection}
\label{sec:data-collection}

    Data collection started with the dataset created by \citet{wang_why_2025}, which contains approximately~1.2 million Jupyter notebooks shared on Kaggle\footnote{Kaggle: \href{https://www.kaggle.com/}{https://www.kaggle.com/}} and GitHub.
    Jupyter notebooks provide an interactive development environment in which developers can run fragments of their code~(code cells) to get quick implementation insights without rerunning the entire code base~\citep{wang_better_2020, pimentel_large-scale_2019, koenzen_code_2020}.
    The evaluation uses their Kaggle notebooks because the public Kaggle API simplifies downloading datasets.
    Kaggle is an online platform for ML competitions, where practitioners compete to train the best-performing model for a specific task.
    To ensure that the analyzed code is fit for our research, notebooks are converted into syntactically correct Python scripts using \texttt{nbconvert}\footnote{\texttt{nbconvert}: \href{https://pypi.org/project/nbconvert/}{https://pypi.org/project/nbconvert/}}
    An overview of our data collection process is shown in \autoref{fig:sampling}, which is detailed in the following sections.
    An overview of the collected dataset's descriptive statistics can be found in \autoref{tab:demographics}.

\input{asset-figure-sampling}

    \paragraph*{Base sample}
        To ensure our evaluation is performed on relevant ML code, a base sample of ML scripts was created.
        \citet{wang_why_2025} describe various notebook-specific interaction patterns that are not possible in Python scripts:~out-of-order execution, unexecuted code cells, and repeated execution of code cells.
        Because our analysis translates notebooks to scripts using \texttt{nbconvert}, which outputs all of the code in the order in which it was written, all entries with cells that were run out of order or not run at all were removed, as this might incorrectly inflate the number of faults detected in the analysis.
        
        Approximately~$21\%$ of the syntactically correct notebooks using Scikit-learn used the \pythoninline{RandomForestClassifier}, making it one of the most common ML algorithms in the dataset.
        Scripts that did not use this algorithm were removed to align with our study's scope.
        Exploratory analysis showed that scripts were not distributed equally across authors~(a power-law distribution).
        Therefore, to mitigate grouping effects that bias our results toward faults made by highly active users, one script was randomly sampled per author.
        Ultimately, this yielded~7,235 ML scripts that apply the \pythoninline{RandomForestClassifier}.

    \paragraph*{Enriched subsample}
        In addition to our base sample, an enriched subsample of ML scripts was created by downloading their respective datasets.
        Because the Kaggle API limits the number of datasets you can download in a period of time, a random sample of scripts was taken.
        
        While we attempted to download all datasets referenced in the scripts, several were inaccessible because they were marked as private.
        Because it is impossible to predict whether a dataset is available before starting the collection process, we iteratively extended our sample until a sufficient number of faults could be analyzed to answer \rqn{1}~(see~\autoref{sec:detection-effectiveness}).
        Ultimately, we sampled~2,165 scripts and downloaded usable datasets for~570~($26.3\%$).

\input{asset-table-demographics.tex}

\subsection{Evaluation Method}

    We perform a mixed-methods empirical evaluation to collect results.
    A quantitative evaluation identifies the detection frequency~(\rqn{1}), runtime performance overhead~(\rqn{2}), and the prevalence of semantic faults~(\rqn{3}).
    A qualitative evaluation determines the precision and relevance of detected faults~(\rqn{1}).
    The following sections describe how these methods were used to answer our research questions.
    
    \subsubsection{Detection Effectiveness}
    \label{sec:evaluation-detection-effectiveness}
        To study the effectiveness of our analysis technique~(\rqn{1}), \pythoninline{dille} was used to analyze the enriched subsample of ML scripts according to the main qualities of static analysis tools~\citep{johnson_why_2013}:~\textit{frequency}, \textit{precision}, and \textit{relevance} of detected faults.

        \paragraph*{Quantitative Analysis} To study detection frequency, the analysis included three levels of dataset availability:~1)~using data properties that are calculated during the analysis, and~2)~using precomputed data properties and~3)~not using data properties at all.
        These configurations were tested by automatically running the analysis on each script in the subsample.
        
        \paragraph*{Qualitative Analysis} To study detection precision and relevance, a manual qualitative evaluation was done on a subset of the faults detected in our quantitative evaluation.
        Because our quantitative evaluation showed that faults are not equally common in ML scripts, we took a stratified random sample to ensure each fault type is represented in our evaluation.
        Based on the time-intensity identified in our trial evaluation, we randomly sampled up to five detected faults per fault type, and fewer if the fault was detected less than five times.
        This yielded a total sample of~37 detected faults for manual evaluation.
        The correctness of detected faults was measured in two ways: 1)~a trinary variable~(correct, incorrect, and uncertain), and 2)~how certain the reviewer was of that classification (5-point Likert scale).

        This evaluation was performed in two stages:~1)~a trial evaluation to determine the evaluation's time intensity, resolve issues in the study design, and identify potential bugs in our tool, and~2)~a main evaluation to generate the results presented in this paper.
        The trial evaluation was performed by the first author~(designer of the analysis technique), and the main evaluation was performed by the first author and second author~(ML expert).
        To ensure an unbiased evaluation, the ML expert was entirely excluded from the design and implementation of \pythoninline{dille}, and their involvement was strictly confined to the manual evaluation phase.
        In addition, relevance was exclusively scored by the ML expert.

    \subsubsection{Runtime Overhead}
        To identify runtime overhead~(\rqn{2}) of our solution, we collected runtime information of the analyses described in \autoref{sec:evaluation-detection-effectiveness}.
        Because we ran our experiments in a shared computation environment,\footnote{\anon{WARA-OPS: \href{https://www.wara-ops.org/}{https://www.wara-ops.org/}}} we repeated each experiment five times to reduce the impact of background noise on our runtime performance measurements.
        Our analysis revealed that approximately~$2\%$ of repeats took over an order of magnitude longer than others.
        Because we were unable to attribute these to any particular analysis, we marked them as background noise and removed them from further analysis.

        Because our solution can replace datasets with precomputed data properties, we remove computational overhead from the analysis.
        This is quantified by calculating the speedup between analyses using datasets and those using precomputed properties.
        Intuitively, the greatest speedups are gained in analyses that are slower by nature, for which we explored the relationship between speedup and~1)~analysis runtime, and~2)~dataset size.
        We investigated their general relationship with speedup by fitting a double-log regression model~\citep{benoit2011linear} because all variables follow power-law distributions.

    \subsubsection{Fault Prevalence}
        Based on the results of \rqn{1}, \pythoninline{dille} can be used to estimate a lower bound of the number of silent semantic faults in ML pipelines that use the \pythoninline{RandomForestClassifier}.
        To do this, all scripts in the base sample were automatically evaluated.
        However, unlike the evaluation for \rqn{1}, this analysis does not leverage datasets, as downloading them would be too time-consuming.
        Although this means that strictly data-dependent semantic faults could not be detected, hyperparameter and structural semantic faults could be detected.
        In addition, the analysis of the enriched subsample provides an initial idea of the prevalence of semantic data faults.
        

%% file: asset-figure-sampling.tex
\begin{figure}[!b]
    \centering
    \includegraphics[width=0.865\columnwidth]{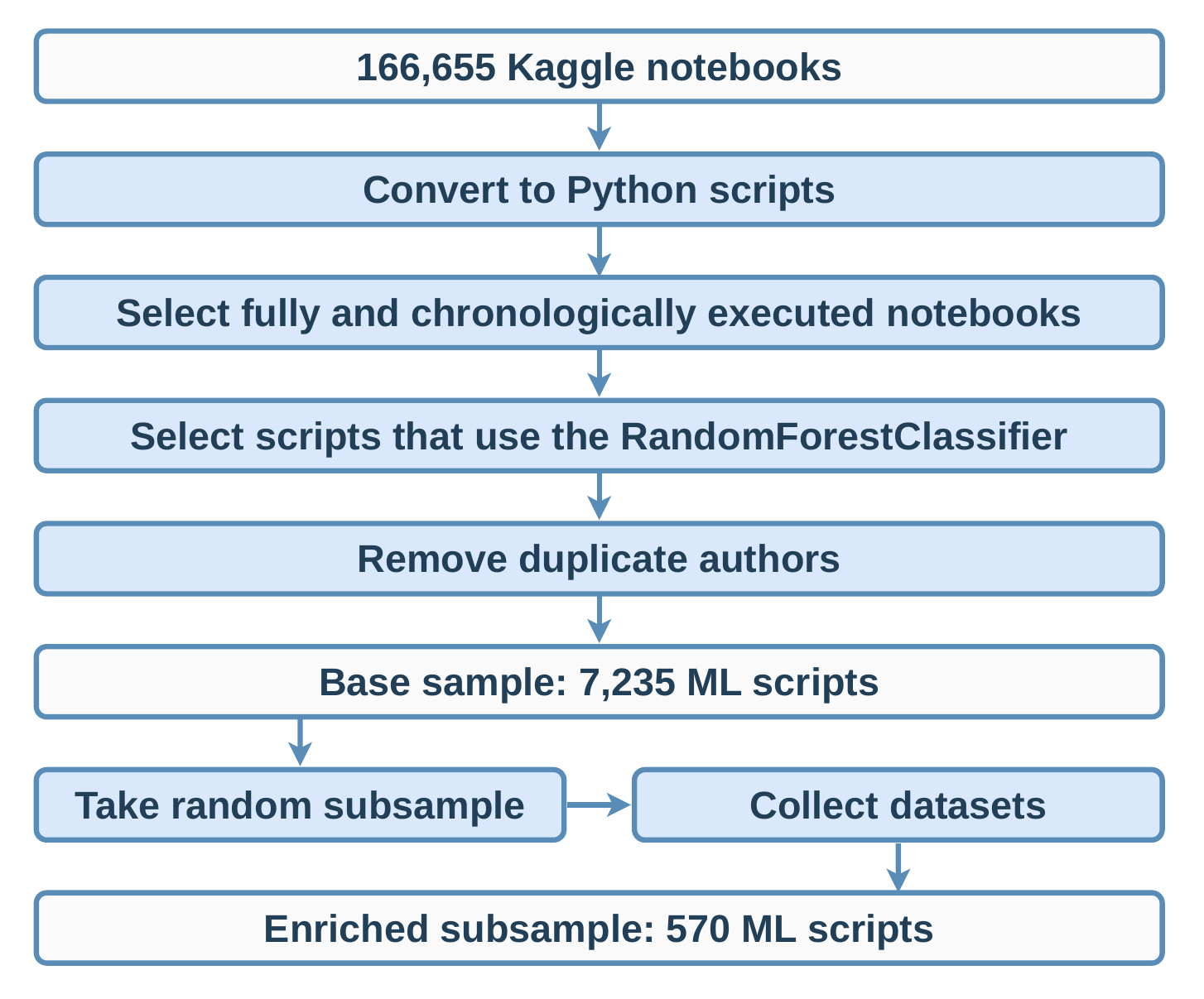}
    \caption{The data collection and filtering process used to create the base sample and enriched subsample.}
    \label{fig:sampling}
\end{figure}

%% file: asset-table-demographics.tex
\begin{table}[!t]
    \caption{Descriptive statistics of the base sample and enriched subsample, showing notebook cell count~(Cells), logical lines of code~(LLoC), dataset size, and Kaggle votes}
    \label{tab:demographics}
    \centering
    \begin{tabularx}{\linewidth}{|l|Y|Y|Y|Y|Y|}
        \hline
        \textbf{Metric} & \textbf{Min.} & \textbf{25th} & \textbf{Med.} & \textbf{75th} & \textbf{Max.} \\
        \hline\hline
        \multicolumn{6}{|l|}{\textbf{Base sample} ($n=$ 7,235)} \\
        \hline
        Cells & 1 & 18 & 30 & 46 & 379\\
        LLoC & 10 & 72 & 120 & 197 & 1774\\
        \hline\hline
        \multicolumn{6}{|l|}{\textbf{Enriched subsample} ($n=$ 570)} \\
        \hline
        Cells & 1 & 20 & 30 & 46 & 245\\
        LLoC & 10.0 & 75.2 & 129.5 & 206.0 & 1189.0\\
        Dataset size & 0 B & 48 KB & 383 KB & 5 MB & 3 GB\\
        Kaggle votes & 0 & 2 & 5 & 12 & 218\\
        \hline
    \end{tabularx}
\end{table}

%% file: section-results.tex
\section{Results}

    Next we discuss the main results of our experimental evaluation separately for each research question.

\input{asset-table-frequency}

\subsection{\rqn{1}: Detection Effectiveness}
\label{sec:detection-effectiveness}

    The effectiveness of our analysis is evaluated along the three main dimensions of static analyzers~\citep{johnson_why_2013}:~the frequency of reported issues, their precision, and relevance in ML development.
    We successfully analyzed a total of~451 out of~570 notebooks~($79\%$) in three configurations of our tool:~1)~using the complete dataset,~2)~using precomputed data properties, and~3)~without access to data.
    We compare our results with \mbox{MLScent}~\citep{shivashankar_mlscent_2025}, a static analysis tool for ML scripts.
    
    \paragraph*{Frequency}
        Our solution detected~595 semantic faults in~264 notebooks~($59\%$).
        An overview of detected faults can be found in \autoref{fig:frequency}.
        Because data is not always available at analysis time, we evaluate our solution with access to precomputed data properties and without access to any data, and compare the results with the analysis with accessible data.
        Analysis with precomputed data properties is~$99.3\%$ consistent with analysis using the complete dataset.
        The results for analysis without data are similar, such that all hyperparameter and structural faults are detected.
        As expected, none of the data faults could be detected as their evaluation depends on data.
        Nevertheless, analysis using no data still detected the majority~($82\%$) of semantic faults.

        In comparison, \mbox{MLScent}~\citep{shivashankar_mlscent_2025} detected~11,230 potential problems in total which were filtered based on related library~(e.g., to exclude problems related to TensorFlow), code style problems~(e.g., to exclude name convention issues), and unrelated issues related to dataset management~(e.g., to exclude issues related to explicitly specifying column names).
        This resulted in only six faults related to inconsistent feature scaling which could potentially overlap with our detected faults. 
        After investigating them one by one, we found that they are all unrelated.
        This highlights that the fault types detected by our solution are totally different from those detected by \mbox{MLScent}.

    \paragraph*{Precision}
        The correctness of our procedure is assessed by manually analyzing~37 detected faults:~15 hyperparameter faults,~13 data faults, and~9 structural faults.
        Our second reviewer (ML expert) marked five faults as uncertain because they did not know the fault type (correlated features and overlapping training samples).
        To identify inter-reviewer agreement about the correctness of detected faults, we calculate Cohen's Kappa~\citep{cohen1960coefficient} on the instances classified by both reviewers.
        Four instances required additional clarification.
        After discussion, this yielded an agreement of~$0.65$~(substantial).
        In general, the reviewers are confident about their classifications, as the mode for both reviewers is 5~(extremely certain), with mean confidences of~$4.5$ and~$4.1$, respectively.
        
        Reviewers disagreed on the correctness of two faults:~an accidentally correct fault~(fault~\#2.11 in the replication package~\citep{meijer_experiment_2026}) and a fault that likely has limited impact~(fault~\#1.26).
        Unresolved entries were discarded, drawing conclusions based on the analyses where both reviewers agreed.
        The reviewers definitively classified~31 faults~($84\%$).
        The two reviewers classified~29 of these faults as correctly identified~($91\%$), where all hyperparameter and data faults were correct, and 6 structural faults were correct~($75\%$).
        Two faults were incorrect, misidentifying unequal data preprocessing~(S1).


    \paragraph*{Relevance}
        Similar to correctness, the relevance of detected faults was evaluated.
        Contrary to identifying precision, relevance was exclusively evaluated by the ML expert.
        Overall, detected faults are moderately/very relevant~(a Likert score of~3.7/5), increasing when the detected faults are correct~(3.9/5).
        These scores are not equal for all correctly detected fault types, as hyperparameter faults are moderately relevant~(3.1/5), data faults are very relevant~(3.8/5), and structural faults are extremely relevant~(4.9/5).

    \paragraph*{Summary}
        Data-informed static analysis can effectively detect random-forest-specific silent semantic faults, identifying silent semantic faults in~$59\%$ of notebooks with~$91\%$ precision.
        Overall, detected faults are moderately/very relevant~(3.7/5), where data and structural faults are particularly relevant~(3.8/5 and~4.9/5, respectively).
        The detected faults are unique, as they were not detected by \mbox{MLScent}~\citep{shivashankar_mlscent_2025}, a state-of-the-art analysis tool for ML code.
        Even without direct data access, our procedure can be used with~$99.3\%$ consistency by computing relevant data properties before the analysis, and with~$81.7\%$ consistency when datasets are completely unavailable.
        This enables using this type of analysis in contexts where data sharing is difficult or limited by confidentiality policies.
    
    
    \takeaway{\rqn{1}}{
        Data-informed static analysis can effectively detect various unique and relevant structural, data, and hyperparameter silent semantic faults, requiring only aggregate data properties rather than complete datasets.
    }

\input{asset-figure-data-speedup}

\subsection{\rqn{2}: Runtime Overhead}
\label{sec:runtime-overhead}

    Because static analysis can be run in different contexts with different runtime overhead requirements~(e.g., \mbox{CI/CD} pipelines or IDEs), we measured the runtime overhead of our technique.
    The median runtime overhead of our solution was~$0.19$ in analyses with access to data, and~$0.16$ in analyses with access to precomputed data properties, where~$92\%$ and~$96\%$ of analyses were completed within one second, respectively.
    This makes our analysis comparable to \mbox{MLScent}, which had a median runtime overhead of~$0.05$ seconds.

    We identified four cases where analyses took more than~5 seconds, which had three root causes:~1)~many lines of code,~2)~many detected faults, and~3)~intensive data calculations in the analysis.
    
    We emphasize the~$111$ data-dependent analyses and the~$54$ analyses that performed more intensive calculations~(e.g., feature correlation).
    An example can be found in \autoref{fig:data-speedup}.
    Using precomputed data properties had a median speedup of~$1.02$ and an average speedup of~$1.5$, with four cases exceeding a speedup of~$10$, and an extreme case of~$105$.
    
    Analysis runtime and dataset size are positively correlated with speedup, increasingly so for data-dependent analyses.
    Analysis runtime is weakly correlated with speedup~($r = 0.36, p < 0.001$), strengthening in data-dependent analyses~($r = 0.50, p < 0.001$), especially intensive calculations~($r = 0.55, p < 0.001$).
    Similarly, dataset size is weakly correlated to speedup~($r = 0.21, p < 0.001$), increasing with greater data-dependency~($r = 0.58, p < 0.001$), especially with intensive data operations~($r = 0.61, p < 0.001$).

    Because visual inspection suggested that analyses with larger datasets and longer runtimes are more strongly correlated to speedup, we split the dataset on the median of the independent variable to perform separate regression analyses on the lower and upper halves.
    While this did not identify any correlation between the independent variables and speedup on the lower halves, the previously reported relationships appeared clearly in the upper halves.
    We highlight the correlation between the larger~$50\%$ of datasets~(i.e., larger than~150~KB) and speedup in data-dependent analyses, shown in \autoref{fig:data-speedup}, which became much stronger~($r = 0.74, p < 0.001$).
    
    \paragraph*{Summary}
        Our solution adds only a small performance overhead, requiring less than one second in $92\%$ to $96\%$ of analyses.
        Overhead is further reduced by using precomputed data properties, achieving a median speedup of~$1.05$ with extreme speedups up to~$105$.
        Analyses with longer runtimes and analyses on notebooks with larger datasets benefit most, showing a statistically significant correlation with speedup in the upper $50\%$.
        This relationship is amplified when analyses rely on heavier data calculations~(e.g., calculating feature correlation).
        This enables its use in time-constrained contexts.
        For example, to provide real-time feedback in IDEs~(which requires sub-second analyses) or in agentic software development~(which requires one/two minutes).

    \takeaway{\rqn{2}}{
        Our data-informed static analysis solution adds sub-second runtime overhead, enabling its use in time-constrained contexts, like IDEs or agentic workflows.
        Precomputing data properties further minimizes analysis overhead, especially on scripts with larger datasets.
    }

\subsection{\rqn{3}: Fault Prevalence}
\label{sec:fault-prevalence}
    
    Using our analysis technique, we can estimate a lower bound of the prevalence of semantic faults in notebooks that use random forest classifiers.
    We analyze~570 notebooks with accessible datasets and~7,235 notebooks without access to datasets to identify the prevalence of semantic faults.
    The results are shown in \autoref{fig:frequency}.

    The evaluation of our enriched subsample detected~595 faults in~264 notebooks~($59\%$).
    The majority~($73\%$) of these regard missing random seeds (H0).
    Although random seeds are essential for replicability, it is a context-dependent requirement.
    Excluding random seed faults, we detected~160 semantic faults in~81 notebooks~($18\%$) where~109 are data faults~($68\%$),~30 are structural faults~($19\%$), and~21 are other hyperparameter faults~($13\%$).

    Our solution successfully analyzed~5,563 of~7,235 notebooks in the base sample~($77\%$).
    It detected~5,358 faults in~2,589 notebooks~($51\%$), of which the majority~($83\%$) related to missing random seeds~(H0).
    Therefore, excluding random seed faults, we detected~902 faults in~573 notebooks~($10\%$), of which~617 are structural faults~($68\%$) and~285 are other hyperparameter faults~($32\%$).
    As the base sample is analyzed without data access, no data faults were detected.

    These results highlight the high prevalence of data and structural faults in ML notebooks.
    Data faults are~$5.2$ times more common than hyperparameter faults in our analysis of the enriched subsample.
    Similarly, structural faults are detected~$1.4$ times more often than hyperparameter faults in the analysis of the enriched sample, and~$2.1$ times more often in the analysis of the base sample.
    Adjusting this for our analysis in \rqn{1}, which identified that~$60\%$ of the detected unequal data preprocessing faults were detected correctly, structural faults remain~$1.3$ times more common.
    This is surprising, as hyperparameter faults are some of the most commonly addressed in state-of-the-art static analysis tools~\citep{ahmed_design_2023, gao_refty_2022, shivashankar_mlscent_2025, reimann_safe-ds_2023, dolcettiPYRAHighlevelLinter2026, hong_investigating_2024}, highlighting a potential mismatch between existing problems and proposed solutions.
    
    \paragraph*{Summary}
        Our solution detects silent semantic faults in~$10\%$ to~$18\%$ of the analyzed notebooks, and in~$51\%$ to~$59\%$ when including faults related to random seeds~(H0).
        Data faults are most common, comprising~$68\%$ of the detected faults in our analysis of the enriched subsample,~$3.6$ times more frequently than hyperparameter faults.
        This is followed by structural faults, containing~$32\%$ of the detected faults, making it approximately $1.3$ times more common than hyperparameter faults.
        These results highlight the high prevalence of silent semantic faults in ML notebooks.

    \takeaway{\rqn{3}}{
        Silent semantic faults are prevalent in ML notebooks that use the random forest classifier.
        Unmanaged randomness accounts for the majority of detected hyperparameter faults.
        Excluding these highlights the frequency of data and structural faults, appearing more frequently than other hyperparameter faults.
    }

%% file: asset-table-frequency.tex
\begin{table}[t]
    \centering
    \caption{The number of detected faults in the three analyses performed on the enriched subsample (with data access, with precomputed properties, without data access), and the analysis performed on the base sample without access to data; percentages under the enriched analysis are the fraction of retained faults compared to the analysis with data access}
    \label{fig:frequency}
    \begin{tabularx}{\columnwidth}{|l|YYY|Y|}
        \hline
        ~ & \multicolumn{3}{c|}{\textit{Enriched analyses}} & \textit{Base analysis} \\
        \hline
        \textbf{ID} & \textbf{Data access} & \textbf{Precomp.} & \textbf{No Data} & \textbf{Descript.} \\
        \hline\hline
        
        \multicolumn{5}{|l|}{\textit{Hyperparameter Faults}} \\
        \hline
        
        \textit{H0} & 435 & 435 (100\%) & 435 (100\%) & 4,456 \\
        \textit{H1} & 3 & 3 (100\%) & 3 (100\%) & 104 \\
        \textit{H2} & 18 & 18 (100\%) & 18 (100\%) & 181 \\
        \hline
        \textbf{Tot.} & \textbf{456} & \textbf{456 (100\%)} & \textbf{456 (100\%)} & \textbf{4,741} \\
        \hline\hline
        
        \multicolumn{5}{|l|}{\textit{Data Faults}}  \\\hline
        \textit{D1} & 28 & 27 (96.4\%) & 0 (0\%) & 0 \\
        \textit{D2} & 1 & 1 (100\%) & 0 (0\%) & 0 \\
        \textit{D3} & 2 & 1 (50.0\%) & 0 (0\%) & 0 \\
        \textit{D4} & 3 & 3 (100\%) & 0 (0\%) & 0 \\
        \textit{D5} & 1 & 1 (100\%) & 0 (0\%) & 0 \\
        \textit{D6} & 74 & 72 (97.3\%) & 0 (0\%) & 0 \\
        \hline
        \textbf{Tot.} & \textbf{109} & \textbf{105~(96.3\%)} & \textbf{0 (0\%)} & \textbf{0} \\
        \hline\hline
        
        \multicolumn{5}{|l|}{\textit{Structural Faults}} \\\hline
        \textit{S1} & 8 & 8 (100\%) & 8 (100\%) & 145 \\
        \textit{S2} & 22 & 22 (100\%) & 22 (100\%) & 472 \\
        \hline
        \textbf{Tot.} & \textbf{30} & \textbf{30 (100\%)} & \textbf{30 (100\%)} & \textbf{617} \\
        \hline\hline
        
        \textbf{Tot.} & \textbf{595} & \textbf{591~(99.3\%)} & \textbf{486~(81.7\%)} & \textbf{5,358} \\
        \hline
    \end{tabularx}
\end{table}

%% file: asset-figure-data-speedup.tex
\begin{figure}[b]
    \centering
    \includegraphics[width=\columnwidth]{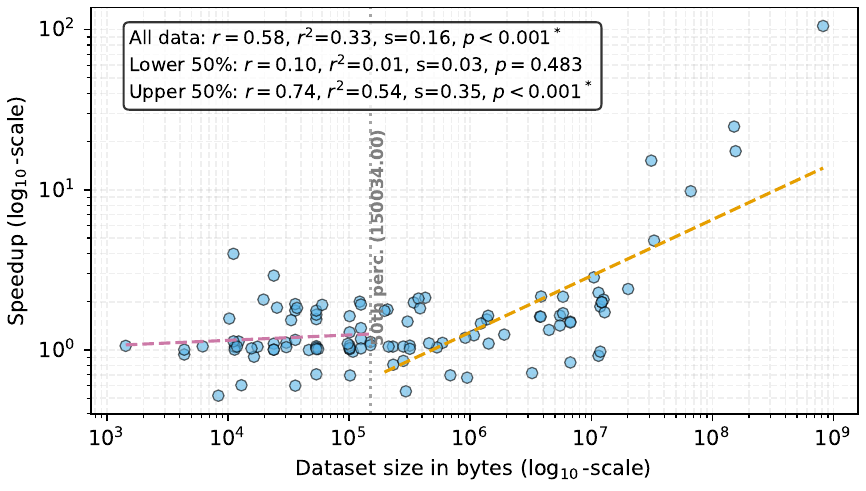}
    \caption{Relationship between dataset size and speedup in data-dependent analyses~($n=111$), highlighting the double-log linear relationships described by their Pearson correlation~$r$, explained variance~$r^2$, regression slope~$s$, and $p$-value.}
    \label{fig:data-speedup}
\end{figure}

%% file: section-threats.tex
\subsection{Threats to Validity}
\label{sec:threats}

\paragraph{Construct validity}
    
    We manually evaluated the precision and relevance of a subset of detected faults based on existing literature~\citep{johnson_why_2013}.
    Because manual analysis is error-prone, each fault was classified by two reviewers.
    We resolved conflicts through consensus and discarded unresolved entries to ensure a conservative evaluation, drawing conclusions based on the analyses on which both reviewers agreed.
    To mitigate confirmation bias in relevance scores, only the relevance scores provided by the machine learning expert were used.
    Because our analysis does not include recall, which indicates the fraction of real faults detected by our procedure, the number of faults that were correctly identified by our procedure provides a lower bound on the number of faults in the analyzed notebooks.

\paragraph{Internal validity}

    We use a random sample of Kaggle notebooks collected by \citet{wang_why_2025} as the starting point of our study.
    Because we convert notebooks to Python scripts, notebook-specific interaction patterns in this dataset may affect the detected faults.
        We could not remove notebooks with repeatedly executed cells because they are overwhelmingly present in the dataset~($98\%$).
    While repeated executions could occlude semantic faults, it is unlikely that we misidentified faults, as we manually validated the correctness of a subset of detected faults.

\paragraph{External validity}

    We evaluate our solution using notebooks from the Kaggle ecosystem.
    While Kaggle notebooks are not industrial machine learning pipelines, they represent diverse data science practices.
    We scope our evaluation of data-informed analysis to random forest classifiers, giving an in-depth evaluation of a specific use case.
    Regardless, the benefits of data-informed static analysis are much broader than that.
    For example, class imbalance~\citep{altalhan_imbalanced_2025} and non-random training and testing data~\citep{kapoor_leakage_2023} are problems for many machine learning algorithms, making our procedure applicable in these contexts as well.
    Our procedure successfully analyzed between $77\%$ and $79\%$ of the scripts in our dataset, where the remainder was not analyzed due to syntactic errors after code canonicalization.
    This creates a risk of survivorship bias if this subset contains significantly different fault types.

        


%% file: section-discussion.tex
\section{Implications}

The results of our evaluation have several implications for the use of static analysis tools for ML code.

\paragraph*{Silent semantic faults are prevalent in random forest classifiers}
    The high prevalence of detected issues~(appearing in~$10\%$ to~$18\%$ of analyzed notebooks) indicates that silent semantic faults are a significant hurdle in ML development.
    Our findings highlight that data and structural faults are particularly frequent, occurring~$1.3$ to~$3.6$ times more often than hyperparameter issues.

\paragraph*{Relevant semantic faults can be detected with high precision}
    Our tool detects silent semantic faults with high precision~($91\%$) that are highly relevant to ML development, especially structural and data faults.
    This shows that shifting the analysis focus to the intersection of code and datasets is essential for improving the quality of ML software.

\paragraph*{Only data properties are needed, not data itself}
    While our analysis technique is data-informed, 
    it only requires access to aggregate data properties to identify silent semantic faults, and not the complete datasets.
    Analyses that use precomputed data properties identify~$99.3\%$ consistent semantic faults compared to analyses with full data access.
    This flexibility allows running our analysis in scenarios where data confidentiality is paramount without losing the ability to detect silent semantic faults.
    If aggregated data properties are computed by a client, the real client data can remain confidential.

\paragraph*{Fault detection can be integrated into time-constrained contexts}
    The sub-second runtime of our solution in~$92\%$ to~$95\%$ of analyses ensures that semantic fault detection is suitable for integration into time-constrained development environments.
    This includes real-time feedback in IDEs~(requiring sub-second analyses), rapid iterative loops in agentic coding frameworks~(allowing up to one/two minutes), and automated checks in \mbox{CI/CD} pipelines~(allowing minutes to hours).
    Our tool can further accelerate these checks via precomputed properties to ensure efficient analysis in the developer workflow even as datasets grow in size and complexity.

\paragraph*{Summary}
    Data-informed static analysis can effectively identify silent semantic faults in ML code that uses the random forest classifier, in particular structural and data faults.
    The high prevalence of identified faults demonstrates the impact on the quality of ML pipelines.
    The procedure is dataset-independent, allowing seamless integration into time-sensitive and security-restricted environments without compromising detection effectiveness.

%% file: section-conclusion.tex
\section{Conclusion}

In this paper, we addressed the \textit{silent semantic faults} in ML scripts that use the popular \pythoninline{RandomForestClassifier}~\citep{pedregosa_scikit-learn_2011, sklearn-rf-doc, breiman_random_2001}.
Silent faults are time- and resource-intensive to detect because they lack apparent symptoms.
We proposed a novel data-informed static analysis technique that leverages API contracts and aggregate data properties to detect unique faults that are not detected by state-of-the-art analysis tools.
We implemented this in an open-source prototype tool called \pythoninline{dille}~\citep{meijer_experiment_2026}.

Our evaluation of~7,235 real-world notebooks yielded several key insights.
First, we demonstrate that silent semantic faults are alarmingly prevalent, affecting~$10\%$ to~$18\%$ of notebooks that use the random forest classifier.
Second, our approach achieves high precision~($91\%$) with sub-second runtime overhead, making it suitable for integration into real-time IDE feedback loops, agentic workflows, and \mbox{CI/CD} pipelines.
Third, our results confirm that using precomputed data properties maintains $99.3\%$ consistency relative to analyses with complete dataset access, which is beneficial in contexts where strict data confidentiality rules apply.

Future work will extend our analysis tool \pythoninline{dille} in three main directions.
First, we will broaden API support to include other ML algorithms such as \pythoninline{LogisticRegression} and optimization frameworks like \pythoninline{GridSearchCV}.
Second, we aim to integrate domain-specific metadata to detect complex silent semantic faults, such as grouping effects and temporal data leakage that aggregate data properties cannot fully capture.
Finally, we will explore the transition from ML scripts (used in experiments) to production-ready ML systems, investigating how our DAG-based pipeline extraction can support the migration of pipelines into larger systems.

%% file: metasection-data-availability.tex
\section*{Data Availability Statement}

The implementation of \pythoninline{dille}, data analysis scripts, and datasets used in this study can be found in our replication package~\citep{meijer_experiment_2026}.

%% file: metasection-credits.tex
\section*{CreDiT Author Contributions}
\anon{\textbf{Willem Meijer:} Conceptualization, Methodology, Software, Validation, Formal analysis, Data Curation, Writing --- Original Draft, Writing --- Review \& Editing, Visualization, Project administration.
\textbf{Louis Ohl:} Formal analysis, Investigation, Validation.
\textbf{Kristian Sandahl and Dániel Varró:} Conceptualization, Methodology, Writing --- Original Draft, Writing --- Review \& Editing, Supervision, Project administration, Funding acquisition.}

%% file: references-anonymized.bib
@misc{meijer_experiment_2026,
  __author = {Meijer, Willem and Sandahl, Kristian and Varró, Dániel},
  author = {Anonymized Authors},
  title = {Replication package for "Are We Lost in the Woods? Detecting Silent Semantic Faults for Random Forest Classifiers with Data-informed Static Analysis"},
  year = 2026,
  publisher = {Zenodo},
  doi = {10.5281/zenodo.19344519},
  url = {https://doi.org/10.5281/zenodo.19344519},
}

@inproceedings{meijer_data-aware_2026,
    author = {Anonymized Authors},
  __publisher = {ACM},
  __author = {Meijer, Willem and Sandahl, Kristian and D{\'a}niel Varr{\'o}},
    title = {Data-aware Static Analysis: Improving Detection of Semantic Faults in Machine Learning Code Using Data Characteristics},
    booktitle = "Proceedings of IEEE/ACM 48th International Conference on Software Engineering: New Ideas and Emerging Results (ICSE-NIER ’26)",
    year = 2026,  
    note = {To appear, an anonymized preprint is added as supplementary material to this submission}

}


%% file: references-doi2bib.bib
@book{oecdbcginsead_adoption_2025,
  title = {The Adoption of Artificial Intelligence in Firms: New Evidence for Policymaking},
  ISBN = {9789264803756},
  url = {http://dx.doi.org/10.1787/f9ef33c3-en},
  DOI = {10.1787/f9ef33c3-en},
  publisher = {OECD Publishing},
  author = {{OECD/BCG/INSEAD}},
  year = {2025},
  month = May 
}

@article{arroyabe_analyzing_2024,
  title = {Analyzing {AI} adoption in {European} {SMEs}: A study of digital capabilities,  innovation,  and external environment},
  volume = {79},
  ISSN = {0160-791X},
  url = {http://dx.doi.org/10.1016/j.techsoc.2024.102733},
  DOI = {10.1016/j.techsoc.2024.102733},
  journal = {Technology in Society},
  publisher = {Elsevier BV},
  author = {Arroyabe,  Marta F. and Arranz,  Carlos F.A. and Fernandez De Arroyabe,  Ignacio and Fernandez de Arroyabe,  Juan Carlos},
  year = {2024},
  month = Dec,
  pages = {102733}
}

@inproceedings{shivashankar_maintainability_2022,
  title = {Maintainability Challenges in {ML}: A Systematic Literature Review},
  url = {http://dx.doi.org/10.1109/SEAA56994.2022.00018},
  DOI = {10.1109/seaa56994.2022.00018},
  booktitle = {2022 48th Euromicro Conference on Software Engineering and Advanced Applications (SEAA)},
  publisher = {IEEE},
  author = {Shivashankar,  Karthik and Martini,  Antonio},
  year = {2022},
  month = Aug,
  pages = {60–67}
}

@inbook{santhanam_quality_2020,
  title = {Quality Management of Machine Learning Systems},
  ISBN = {9783030621445},
  ISSN = {1865-0937},
  url = {http://dx.doi.org/10.1007/978-3-030-62144-5_1},
  DOI = {10.1007/978-3-030-62144-5_1},
  booktitle = {Engineering Dependable and Secure Machine Learning Systems},
  publisher = {Springer International Publishing},
  author = {Santhanam,  P.},
  year = {2020},
  pages = {1–13}
}

@inproceedings{bogner_characterizing_2021,
  title = {Characterizing Technical Debt and Antipatterns in {AI}-Based Systems: A Systematic Mapping Study},
  url = {http://dx.doi.org/10.1109/TechDebt52882.2021.00016},
  DOI = {10.1109/techdebt52882.2021.00016},
  booktitle = {2021 IEEE/ACM International Conference on Technical Debt (TechDebt)},
  publisher = {IEEE},
  author = {Bogner,  Justus and Verdecchia,  Roberto and Gerostathopoulos,  Ilias},
  year = {2021},
  month = May,
  pages = {64–73}
}

@article{cote_quality_2024,
  title = {Quality issues in machine learning software systems},
  volume = {29},
  ISSN = {1573-7616},
  url = {http://dx.doi.org/10.1007/s10664-024-10536-7},
  DOI = {10.1007/s10664-024-10536-7},
  number = {6},
  journal = {Empirical Software Engineering},
  publisher = {Springer Science and Business Media LLC},
  author = {C\^oté,  Pierre-Olivier and Nikanjam,  Amin and Bouchoucha,  Rached and Basta,  Ilan and Abidi,  Mouna and Khomh,  Foutse},
  year = {2024},
  month = Sept 
}

@article{serban_software_2024,
  title = {Software engineering practices for machine learning — Adoption,  effects,  and team assessment},
  volume = {209},
  ISSN = {0164-1212},
  url = {http://dx.doi.org/10.1016/j.jss.2023.111907},
  DOI = {10.1016/j.jss.2023.111907},
  journal = {Journal of Systems and Software},
  publisher = {Elsevier BV},
  author = {Serban,  Alex and van der Blom,  Koen and Hoos,  Holger and Visser,  Joost},
  year = {2024},
  month = Mar,
  pages = {111907}
}

@article{whang_data_2023,
  title = {Data collection and quality challenges in deep learning: a data-centric {AI} perspective},
  volume = {32},
  ISSN = {0949-877X},
  url = {http://dx.doi.org/10.1007/s00778-022-00775-9},
  DOI = {10.1007/s00778-022-00775-9},
  number = {4},
  journal = {The VLDB Journal},
  publisher = {Springer Science and Business Media LLC},
  author = {Whang,  Steven Euijong and Roh,  Yuji and Song,  Hwanjun and Lee,  Jae-Gil},
  year = {2023},
  month = Jan,
  pages = {791–813}
}

@article{kumarOpportunitiesChallengesDataCentric2024,
  title = {Opportunities and Challenges in Data-Centric {AI}},
  volume = {12},
  ISSN = {2169-3536},
  url = {http://dx.doi.org/10.1109/ACCESS.2024.3369417},
  DOI = {10.1109/access.2024.3369417},
  journal = {IEEE Access},
  publisher = {Institute of Electrical and Electronics Engineers (IEEE)},
  author = {Kumar,  Sushant and Datta,  Sumit and Singh,  Vishakha and Singh,  Sanjay Kumar and Sharma,  Ritesh},
  year = {2024},
  pages = {33173–33189}
}

@article{albelali_testing_2025,
  title = {Testing Machine Learning and Deep Learning Systems: Achievements and Challenges},
  volume = {50},
  ISSN = {2191-4281},
  url = {http://dx.doi.org/10.1007/s13369-025-10276-w},
  DOI = {10.1007/s13369-025-10276-w},
  number = {15},
  journal = {Arabian Journal for Science and Engineering},
  publisher = {Springer Science and Business Media LLC},
  author = {Albelali,  Salma and Ahmed,  Moataz},
  year = {2025},
  month = June,
  pages = {11433–11484}
}

@article{nazir_architecting_2024,
  title = {Architecting {ML}-enabled systems: Challenges,  best practices, and design decisions},
  volume = {207},
  ISSN = {0164-1212},
  url = {http://dx.doi.org/10.1016/j.jss.2023.111860},
  DOI = {10.1016/j.jss.2023.111860},
  journal = {Journal of Systems and Software},
  publisher = {Elsevier BV},
  author = {Nazir,  Roger and Bucaioni,  Alessio and Pelliccione,  Patrizio},
  year = {2024},
  month = Jan,
  pages = {111860}
}

@article{bucaioni_checklist_2025,
  title = {A checklist of quality concerns for architecting {ML}-intensive systems},
  volume = {231},
  ISSN = {0164-1212},
  url = {http://dx.doi.org/10.1016/j.jss.2025.112612},
  DOI = {10.1016/j.jss.2025.112612},
  journal = {Journal of Systems and Software},
  publisher = {Elsevier BV},
  author = {Bucaioni,  Alessio and Kazman,  Rick and Pelliccione,  Patrizio},
  year = {2026},
  month = Jan,
  pages = {112612}
}

@inproceedings{humbatova_taxonomy_2020,
  series = {ICSE ’20},
  title = {Taxonomy of real faults in deep learning systems},
  url = {http://dx.doi.org/10.1145/3377811.3380395},
  DOI = {10.1145/3377811.3380395},
  booktitle = {Proceedings of the ACM/IEEE 42nd International Conference on Software Engineering},
  publisher = {ACM},
  author = {Humbatova,  Nargiz and Jahangirova,  Gunel and Bavota,  Gabriele and Riccio,  Vincenzo and Stocco,  Andrea and Tonella,  Paolo},
  year = {2020},
  month = June,
  pages = {1110–1121},
  collection = {ICSE ’20}
}

@article{de_santana_bug_2024,
  title = {Bug Analysis in {Jupyter} Notebook Projects: An Empirical Study},
  volume = {33},
  ISSN = {1557-7392},
  url = {http://dx.doi.org/10.1145/3641539},
  DOI = {10.1145/3641539},
  number = {4},
  journal = {ACM Transactions on Software Engineering and Methodology},
  publisher = {Association for Computing Machinery (ACM)},
  author = {De Santana,  Taijara Loiola and Neto,  Paulo Anselmo Da Mota Silveira and De Almeida,  Eduardo Santana and Ahmed,  Iftekhar},
  year = {2024},
  month = Apr,
  pages = {1–34}
}

@article{khairunnesa_what_2023,
  title = {What kinds of contracts do {ML} {APIs} need?},
  volume = {28},
  ISSN = {1573-7616},
  url = {http://dx.doi.org/10.1007/s10664-023-10320-z},
  DOI = {10.1007/s10664-023-10320-z},
  number = {6},
  journal = {Empirical Software Engineering},
  publisher = {Springer Science and Business Media LLC},
  author = {Khairunnesa,  Samantha Syeda and Ahmed,  Shibbir and Imtiaz,  Sayem Mohammad and Rajan,  Hridesh and Leavens,  Gary T.},
  year = {2023},
  month = Oct 
}

@article{lai_comparative_2024,
  title = {Comparative analysis of real issues in open-source machine learning projects},
  volume = {29},
  ISSN = {1573-7616},
  url = {http://dx.doi.org/10.1007/s10664-024-10467-3},
  DOI = {10.1007/s10664-024-10467-3},
  number = {3},
  journal = {Empirical Software Engineering},
  publisher = {Springer Science and Business Media LLC},
  author = {Lai,  Tuan Dung and Simmons,  Anj and Barnett,  Scott and Schneider,  Jean-Guy and Vasa,  Rajesh},
  year = {2024},
  month = May 
}

@article{morovati_bug_2024,
  title = {Bug characterization in machine learning-based systems},
  volume = {29},
  ISSN = {1573-7616},
  url = {http://dx.doi.org/10.1007/s10664-023-10400-0},
  DOI = {10.1007/s10664-023-10400-0},
  number = {1},
  journal = {Empirical Software Engineering},
  publisher = {Springer Science and Business Media LLC},
  author = {Morovati,  Mohammad Mehdi and Nikanjam,  Amin and Tambon,  Florian and Khomh,  Foutse and Jiang,  Zhen Ming},
  year = {2023},
  month = Dec 
}

@inproceedings{gao_refty_2022,
  series = {ICSE ’22},
  title = {{Refty}: refinement types for valid deep learning models},
  url = {http://dx.doi.org/10.1145/3510003.3510077},
  DOI = {10.1145/3510003.3510077},
  booktitle = {Proceedings of the 44th International Conference on Software Engineering},
  publisher = {ACM},
  author = {Gao,  Yanjie and Li,  Zhengxian and Lin,  Haoxiang and Zhang,  Hongyu and Wu,  Ming and Yang,  Mao},
  year = {2022},
  month = May,
  pages = {1843–1855},
  collection = {ICSE ’22}
}

@inproceedings{reimann_safe-ds_2023,
  title = {{Safe-DS}: A Domain Specific Language to Make Data Science Safe},
  url = {http://dx.doi.org/10.1109/ICSE-NIER58687.2023.00019},
  DOI = {10.1109/icse-nier58687.2023.00019},
  booktitle = {2023 IEEE/ACM 45th International Conference on Software Engineering: New Ideas and Emerging Results (ICSE-NIER)},
  publisher = {IEEE},
  author = {Reimann,  Lars and Kniesel-W\"{u}nsche,  G\"{u}nter},
  year = {2023},
  month = May,
  pages = {72–77}
}

@inproceedings{shivashankar_mlscent_2025,
  title = {{MLScent}: A Tool for Anti-Pattern Detection in {ML} Projects},
  url = {http://dx.doi.org/10.1109/CAIN66642.2025.00026},
  DOI = {10.1109/cain66642.2025.00026},
  booktitle = {2025 IEEE/ACM 4th International Conference on AI Engineering – Software Engineering for AI (CAIN)},
  publisher = {IEEE},
  author = {Shivashankar,  Karthik and Martini,  Antonio},
  year = {2025},
  month = Apr,
  pages = {150–160}
}

@inproceedings{ahmed_design_2023,
  series = {ESEC/FSE ’23},
  title = {Design by Contract for Deep Learning {APIs}},
  url = {http://dx.doi.org/10.1145/3611643.3616247},
  DOI = {10.1145/3611643.3616247},
  booktitle = {Proceedings of the 31st ACM Joint European Software Engineering Conference and Symposium on the Foundations of Software Engineering},
  publisher = {ACM},
  author = {Ahmed,  Shibbir and Imtiaz,  Sayem Mohammad and Khairunnesa,  Samantha Syeda and Cruz,  Breno Dantas and Rajan,  Hridesh},
  year = {2023},
  month = Nov,
  pages = {94–106},
  collection = {ESEC/FSE ’23}
}

@inproceedings{turcotte_fault_2025,
  title = {The Fault in our Stats},
  url = {http://dx.doi.org/10.1109/ASE63991.2025.00205},
  DOI = {10.1109/ase63991.2025.00205},
  booktitle = {2025 40th IEEE/ACM International Conference on Automated Software Engineering (ASE)},
  publisher = {IEEE},
  author = {Turcotte,  Alexi and Mehta,  Neev Nirav},
  year = {2025},
  month = Nov,
  pages = {2491–2503}
}

@article{breiman_random_2001,
  title = {Random Forests},
  volume = {45},
  ISSN = {1573-0565},
  url = {http://dx.doi.org/10.1023/A:1010933404324},
  DOI = {10.1023/a:1010933404324},
  number = {1},
  journal = {Machine Learning},
  publisher = {Springer Science and Business Media LLC},
  author = {Breiman,  Leo},
  year = {2001},
  month = Oct,
  pages = {5–32}
}

@article{fox_assessing_2017,
  title = {Assessing the accuracy and stability of variable selection methods for random forest modeling in ecology},
  volume = {189},
  ISSN = {1573-2959},
  url = {http://dx.doi.org/10.1007/s10661-017-6025-0},
  DOI = {10.1007/s10661-017-6025-0},
  number = {7},
  journal = {Environmental Monitoring and Assessment},
  publisher = {Springer Science and Business Media LLC},
  author = {Fox,  Eric W. and Hill,  Ryan A. and Leibowitz,  Scott G. and Olsen,  Anthony R. and Thornbrugh,  Darren J. and Weber,  Marc H.},
  year = {2017},
  month = June 
}

@article{cappelli_random_2024,
  title = {Random Forest and Feature Importance Measures for Discriminating the Most Influential Environmental Factors in Predicting Cardiovascular and Respiratory Diseases},
  volume = {21},
  ISSN = {1660-4601},
  url = {http://dx.doi.org/10.3390/ijerph21070867},
  DOI = {10.3390/ijerph21070867},
  number = {7},
  journal = {International Journal of Environmental Research and Public Health},
  publisher = {MDPI AG},
  author = {Cappelli,  Francesco and Castronuovo,  Gianfranco and Grimaldi,  Salvatore and Telesca,  Vito},
  year = {2024},
  month = July,
  pages = {867}
}

@inproceedings{grinsztajn_why_2022,
 author = {Grinsztajn, Leo and Oyallon, Edouard and Varoquaux, Gael},
 booktitle = {Advances in Neural Information Processing Systems},
 editor = {S. Koyejo and S. Mohamed and A. Agarwal and D. Belgrave and K. Cho and A. Oh},
 pages = {507--520},
 publisher = {Curran Associates, Inc.},
 title = {Why do tree-based models still outperform deep learning on typical tabular data?},
 url = {https://proceedings.neurips.cc/paper_files/paper/2022/file/0378c7692da36807bdec87ab043cdadc-Paper-Datasets_and_Benchmarks.pdf},
 volume = {35},
 year = {2022}
}

@misc{sklearn-rf-doc,
  author       = {{Scikit-learn}},
  title        = {sklearn.ensemble.{R}andom{F}orest{C}lassifier --- scikit-learn 1.8 documentation},
  howpublished = {\url{https://scikit-learn.org/1.8/modules/generated/sklearn.ensemble.RandomForestClassifier.html}},
  year         = {2025},
}

@misc{sklearn-dt-doc,
  author       = {{Scikit-learn}},
  title        = {sklearn.tree.{D}ecision{T}ree{C}lassifier --- scikit-learn 1.8 documentation},
  howpublished = {\url{https://scikit-learn.org/1.8/modules/generated/sklearn.tree.DecisionTreeClassifier.html}},
  year         = {2025},
}

@article{probst_hyperparameters_2019,
  title = {Hyperparameters and tuning strategies for random forest},
  volume = {9},
  ISSN = {1942-4795},
  url = {http://dx.doi.org/10.1002/widm.1301},
  DOI = {10.1002/widm.1301},
  number = {3},
  journal = {WIREs Data Mining and Knowledge Discovery},
  publisher = {Wiley},
  author = {Probst,  Philipp and Wright,  Marvin N. and Boulesteix,  Anne‐Laure},
  year = {2019},
  month = Jan 
}

@inbook{weiss_mining_2010,
  title = {Mining with Rare Cases},
  ISBN = {9780387098234},
  url = {http://dx.doi.org/10.1007/978-0-387-09823-4_38},
  DOI = {10.1007/978-0-387-09823-4_38},
  booktitle = {Data Mining and Knowledge Discovery Handbook},
  publisher = {Springer US},
  author = {Weiss,  Gary M.},
  year = {2009},
  pages = {747–757}
}

@inbook{cieslak_learning_2008,
  title = {Learning Decision Trees for Unbalanced Data},
  ISBN = {9783540874799},
  ISSN = {1611-3349},
  url = {http://dx.doi.org/10.1007/978-3-540-87479-9_34},
  DOI = {10.1007/978-3-540-87479-9_34},
  booktitle = {Machine Learning and Knowledge Discovery in Databases},
  publisher = {Springer Berlin Heidelberg},
  author = {Cieslak,  David A. and Chawla,  Nitesh V.},
  pages = {241–256}
}

@article{chen_using_2004,
  title={Using random forest to learn imbalanced data},
  author={Chen, Chao and Liaw, Andy and Breiman, Leo},
  journal={University of California, Berkeley},
  volume={110},
  number={1-12},
  pages={24},
  year={2004}
}

@article{nicodemus_behaviour_2010,
  title = {The behaviour of random forest permutation-based variable importance measures under predictor correlation},
  volume = {11},
  ISSN = {1471-2105},
  url = {http://dx.doi.org/10.1186/1471-2105-11-110},
  DOI = {10.1186/1471-2105-11-110},
  number = {1},
  journal = {BMC Bioinformatics},
  publisher = {Springer Science and Business Media LLC},
  author = {Nicodemus,  Kristin K and Malley,  James D and Strobl,  Carolin and Ziegler,  Andreas},
  year = {2010},
  month = Feb 
}

@article{efron_prediction_2020,
  title = {Prediction, Estimation, and Attribution},
  volume = {88},
  ISSN = {1751-5823},
  url = {http://dx.doi.org/10.1111/insr.12409},
  DOI = {10.1111/insr.12409},
  number = {S1},
  journal = {International Statistical Review},
  publisher = {Wiley},
  author = {Efron,  Bradley},
  year = {2020},
  month = Dec 
}

@misc{sklearn-le-doc,
  author       = {{Scikit-learn}},
  title        = {sklearn.preprocessing.{L}abel{E}ncoder --- scikit-learn 1.8 documentation},
  howpublished = {\url{https://scikit-learn.org/1.8/modules/generated/sklearn.preprocessing.LabelEncoder.html}},
  year         = {2025},
}

@book{breiman_classification_2017,
  title = {Classification And Regression Trees},
  ISBN = {9781315139470},
  url = {http://dx.doi.org/10.1201/9781315139470},
  DOI = {10.1201/9781315139470},
  publisher = {Routledge},
  author = {Breiman,  Leo and Friedman,  Jerome H. and Olshen,  Richard A. and Stone,  Charles J.},
  year = {2017},
  month = Oct 
}

@misc{wiryaseputra2023health,
  title        = {{K}aggle Notebook --- Health Insurance Prediction 94\%},
  year         = {2023},
  author       = {Michael Wiryaseputra},
  howpublished = {Kaggle. \url{https://www.kaggle.com/code/michaelwiryaseputra/health-insurance-prediction-94}}
}

@standard{iso24765,
  title        = {Systems and Software Engineering -- Vocabulary},
  author       = {{ISO/IEC/IEEE}},
  number       = {24765:2017},
  year         = {2017},
  publisher    = {International Organization for Standardization},
  address      = {Geneva, Switzerland}
}

@article{pedregosa_scikit-learn_2011,
  title={{Scikit-learn}: Machine learning in {Python}},
  author={Pedregosa, Fabian and Varoquaux, Ga{\"e}l and Gramfort, Alexandre and Michel, Vincent and Thirion, Bertrand and Grisel, Olivier and Blondel, Mathieu and Prettenhofer, Peter and Weiss, Ron and Dubourg, Vincent and others},
  journal={the Journal of machine Learning research},
  volume={12},
  pages={2825--2830},
  year={2011},
  publisher={JMLR. org}
}

@article{dolcettiPYRAHighlevelLinter2026,
  title = {{PYRA}: A high-level linter for data science software},
  volume = {337},
  ISSN = {0950-7051},
  url = {http://dx.doi.org/10.1016/j.knosys.2026.115412},
  DOI = {10.1016/j.knosys.2026.115412},
  journal = {Knowledge-Based Systems},
  publisher = {Elsevier BV},
  author = {Dolcetti,  Greta and Arceri,  Vincenzo and Mensi,  Antonella and Zaffanella,  Enea and Urban,  Caterina and Cortesi,  Agostino},
  year = {2026},
  month = Mar,
  pages = {115412}
}

@article{turcotte_expressing_2025,
  title = {Expressing and Checking Statistical Assumptions},
  volume = {2},
  ISSN = {2994-970X},
  url = {http://dx.doi.org/10.1145/3729391},
  DOI = {10.1145/3729391},
  number = {FSE},
  journal = {Proceedings of the ACM on Software Engineering},
  publisher = {Association for Computing Machinery (ACM)},
  author = {Turcotte,  Alexi and Wu,  Zheyuan},
  year = {2025},
  month = June,
  pages = {2735–2758}
}

@inproceedings{hong_investigating_2024,
  title = {Investigating and Detecting Silent Bugs in {PyTorch} Programs},
  url = {http://dx.doi.org/10.1109/SANER60148.2024.00035},
  DOI = {10.1109/saner60148.2024.00035},
  booktitle = {2024 IEEE International Conference on Software Analysis,  Evolution and Reengineering (SANER)},
  publisher = {IEEE},
  author = {Hong,  Shuo and Sun,  Hailong and Gao,  Xiang and Tan,  Shin Hwei},
  year = {2024},
  month = Mar,
  pages = {272–283}
}

@article{chen_towards_2025,
  title = {Towards Understanding Fine-Grained Programming Mistakes and Fixing Patterns in Data Science},
  volume = {2},
  ISSN = {2994-970X},
  url = {http://dx.doi.org/10.1145/3729352},
  DOI = {10.1145/3729352},
  number = {FSE},
  journal = {Proceedings of the ACM on Software Engineering},
  publisher = {Association for Computing Machinery (ACM)},
  author = {Chen,  Wei-Hao and Cheoh,  Jia Lin and Keim,  Manthan and Brunswicker,  Sabine and Zhang,  Tianyi},
  year = {2025},
  month = June,
  pages = {1824–1846}
}

@article{wang_why_2025,
  title = {Why Do Machine Learning Notebooks Crash? An Empirical Study on Public {Python} {Jupyter} Notebooks},
  volume = {51},
  ISSN = {2326-3881},
  url = {http://dx.doi.org/10.1109/TSE.2025.3574500},
  DOI = {10.1109/tse.2025.3574500},
  number = {7},
  journal = {IEEE Transactions on Software Engineering},
  publisher = {Institute of Electrical and Electronics Engineers (IEEE)},
  author = {Wang,  Yiran and Meijer,  Willem and López,  José Antonio Hernández and Nilsson,  Ulf and Varró,  Dániel},
  year = {2025},
  month = July,
  pages = {2181–2196}
}

@inproceedings{zhang_empirical_2018,
  series = {ISSTA ’18},
  title = {An empirical study on {TensorFlow} program bugs},
  url = {http://dx.doi.org/10.1145/3213846.3213866},
  DOI = {10.1145/3213846.3213866},
  booktitle = {Proceedings of the 27th ACM SIGSOFT International Symposium on Software Testing and Analysis},
  publisher = {ACM},
  author = {Zhang,  Yuhao and Chen,  Yifan and Cheung,  Shing-Chi and Xiong,  Yingfei and Zhang,  Lu},
  year = {2018},
  month = July,
  pages = {129–140},
  collection = {ISSTA ’18}
}

@inproceedings{islam_comprehensive_2019,
  series = {ESEC/FSE ’19},
  title = {A comprehensive study on deep learning bug characteristics},
  url = {http://dx.doi.org/10.1145/3338906.3338955},
  DOI = {10.1145/3338906.3338955},
  booktitle = {Proceedings of the 2019 27th ACM Joint Meeting on European Software Engineering Conference and Symposium on the Foundations of Software Engineering},
  publisher = {ACM},
  author = {Islam,  Md Johirul and Nguyen,  Giang and Pan,  Rangeet and Rajan,  Hridesh},
  year = {2019},
  month = Aug,
  pages = {510–520},
  collection = {ESEC/FSE ’19}
}

@inproceedings{zhang_empirical_2020,
  series = {ICSE ’20},
  title = {An empirical study on program failures of deep learning jobs},
  url = {http://dx.doi.org/10.1145/3377811.3380362},
  DOI = {10.1145/3377811.3380362},
  booktitle = {Proceedings of the ACM/IEEE 42nd International Conference on Software Engineering},
  publisher = {ACM},
  author = {Zhang,  Ru and Xiao,  Wencong and Zhang,  Hongyu and Liu,  Yu and Lin,  Haoxiang and Yang,  Mao},
  year = {2020},
  month = June,
  pages = {1159–1170},
  collection = {ICSE ’20}
}

@inproceedings{johnson_why_2013,
  title = {Why don’t software developers use static analysis tools to find bugs?},
  url = {http://dx.doi.org/10.1109/ICSE.2013.6606613},
  DOI = {10.1109/icse.2013.6606613},
  booktitle = {2013 35th International Conference on Software Engineering (ICSE)},
  publisher = {IEEE},
  author = {Johnson,  Brittany and Song,  Yoonki and Murphy-Hill,  Emerson and Bowdidge,  Robert},
  year = {2013},
  month = May,
  pages = {672–681}
}

@inproceedings{wang_better_2020,
  series = {ICSE ’20},
  title = {Better code,  better sharing: on the need of analyzing {Jupyter} notebooks},
  url = {http://dx.doi.org/10.1145/3377816.3381724},
  DOI = {10.1145/3377816.3381724},
  booktitle = {Proceedings of the ACM/IEEE 42nd International Conference on Software Engineering: New Ideas and Emerging Results},
  publisher = {ACM},
  author = {Wang,  Jiawei and Li,  Li and Zeller,  Andreas},
  year = {2020},
  month = June,
  pages = {53–56},
  collection = {ICSE ’20}
}

@inproceedings{pimentel_large-scale_2019,
  title = {A Large-Scale Study About Quality and Reproducibility of {Jupyter} Notebooks},
  url = {http://dx.doi.org/10.1109/MSR.2019.00077},
  DOI = {10.1109/msr.2019.00077},
  booktitle = {2019 IEEE/ACM 16th International Conference on Mining Software Repositories (MSR)},
  publisher = {IEEE},
  author = {Pimentel,  Joao Felipe and Murta,  Leonardo and Braganholo,  Vanessa and Freire,  Juliana},
  year = {2019},
  month = May,
  pages = {507–517}
}

@inproceedings{koenzen_code_2020,
  title = {Code Duplication and Reuse in {Jupyter} Notebooks},
  url = {http://dx.doi.org/10.1109/VL/HCC50065.2020.9127202},
  DOI = {10.1109/vl/hcc50065.2020.9127202},
  booktitle = {2020 IEEE Symposium on Visual Languages and Human-Centric Computing (VL/HCC)},
  publisher = {IEEE},
  author = {Koenzen,  Andreas P. and Ernst,  Neil A. and Storey,  Margaret-Anne D.},
  year = {2020},
  month = Aug,
  pages = {1–9}
}

@article{benoit2011linear,
  title={Linear regression models with logarithmic transformations},
  author={Benoit, Kenneth},
  journal={London School of Economics, London},
  volume={22},
  number={1},
  pages={23--36},
  year={2011}
}

@article{cohen1960coefficient,
  title={A coefficient of agreement for nominal scales},
  author={Cohen, Jacob},
  journal={Educational and Psychological Measurement},
  volume={20},
  number={1},
  pages={37--46},
  year={1960},
  publisher={Sage Publications}
}

@article{altalhan_imbalanced_2025,
  title = {Imbalanced Data Problem in Machine Learning: A Review},
  volume = {13},
  ISSN = {2169-3536},
  url = {http://dx.doi.org/10.1109/ACCESS.2025.3531662},
  DOI = {10.1109/access.2025.3531662},
  journal = {IEEE Access},
  publisher = {Institute of Electrical and Electronics Engineers (IEEE)},
  author = {Altalhan,  Manahel and Algarni,  Abdulmohsen and Turki-Hadj Alouane,  Monia},
  year = {2025},
  pages = {13686–13699}
}

@article{kapoor_leakage_2023,
  title = {Leakage and the reproducibility crisis in machine-learning-based science},
  volume = {4},
  ISSN = {2666-3899},
  url = {http://dx.doi.org/10.1016/j.patter.2023.100804},
  DOI = {10.1016/j.patter.2023.100804},
  number = {9},
  journal = {Patterns},
  publisher = {Elsevier BV},
  author = {Kapoor,  Sayash and Narayanan,  Arvind},
  year = {2023},
  month = Sept,
  pages = {100804}
}
